\documentclass{article}

\usepackage{arxiv}

\usepackage[utf8]{inputenc} 
\usepackage[T1]{fontenc}    
\usepackage{url}            
\usepackage{booktabs}       
\usepackage{amsfonts}       
\usepackage{nicefrac}       
\usepackage{microtype}      

\usepackage{natbib}
\usepackage{doi}
\usepackage{amsmath,amssymb,bm}
\usepackage{graphicx}       

\graphicspath{{figs/}}     

\usepackage{listings}
\DeclareMathOperator{\sech}{sech}

\title{Blending Bathymetry: Combination of image-derived parametric approximations and celerity data sets for nearshore bathymetry estimation}

\author{
  Jonghyun Lee \\
  Department of Civil and Environmental Engineering \& Water Resources Research Center\\
  University of Hawaii at Manoa \\
  Honolulu, HI USA\\
  \texttt{jonghyun.harry.lee@hawaii.edu} \\
   \And
  Katherine DeVore, Tyler Hesser \\
  Coastal and Hydraulics Laboratory \\
  U.S. Army Engineer Research and Development Center\\
  Vicksburg, MS USA\\
  \texttt{\{Katherine.R.DeVore, Tyler.Hesser\}@erdc.dren.mil} \\
  \AND
  A. Spicer Bak, Katherine Brodie, Brittany Bruder  \\
  Coastal and Hydraulics Laboratory \\  
  U.S. Army Engineer Research and Development Center\\
  Duck, NC USA\\
  \texttt{\{Spicer.Bak, Katherine.L.Brodie. Brittany.L.Bruder\}@erdc.dren.mil} \\
  \And
  Matthew Farthing\\
  Coastal and Hydraulics Laboratory \\
  U.S. Army Engineer Research and Development Center\\
  Vicksburg, MS USA\\
  \texttt{Matthew.W.Farthing@erdc.dren.mil} \\
}

\renewcommand{\shorttitle}{Blending Bathymetry for Real-time Bathymetry Estimation}

\begin{document}
\maketitle

\begin{abstract}
Estimation of nearshore bathymetry is important for accurate prediction of nearshore wave conditions. However, direct bathymetry data collection is expensive and time-consuming while accurate airborne lidar-based survey is limited by breaking waves and decreased light penetration affected by water turbidity. Instead, tower-based platforms or Unmanned Aircraft System (UAS) can provide indirect video-based observations such as time-series (or videos) and time-averaged (timex) or variance enhanced (var) images. The time-series imagery can provide wave celerity information for bathymetry estimation through the well-known dispersion relationship, for example the cBathy algorithm, or physics-based models. However, wave celerities and associated inverted water depths are sensitive to noise during image collection and processing stages or may not even be available over the entire area of interest. Timex or Var images can be used to identify persistent regions of wave breaking (for example over the sand bar and at the shoreline) so that one can create bathymetry profiles using simplified approximations based on parametric forms. However, the accuracy of this approach highly depends on the assumption of the chosen parametric form as well as the accuracy of detecting sandbars and shoreline.

In this work, we propose a rapid and improved bathymetry estimation method that takes advantage of image-derived wave celerity from cBathy and a first-order bathymetry estimate from Parameter Beach Tool (PBT), software that fits parameterized sandbar and slope forms to the nearshore imagery. Two different sources of the data, PBT and wave celerity, are combined or blended optimally based on their assumed accuracy in a statistical (i.e., Bayesian) framework. The PBT-derived bathymetry serves as ``prior'' coarse-scale background information and then is updated and corrected with the cBathy-derived wave data through the dispersion relationship, which results in a better bathymetry estimate that is consistent with imagery-based wave data. To illustrate the accuracy of our proposed method, imagery data sets collected in 2017 at the US Army Engineer Research and Development Center's (ERDC) Field Research Facility (FRF) in Duck, North Carolina under different weather and wave height conditions are tested. Estimated bathymetry profiles are remarkably close to the direct survey data due to the optimal fusion of two data sets. The computational time for the estimation from PBT-based bathymetry and CBathy-derived wave celerity is only about five minutes on a free Google Cloud node with one CPU core. These promising results indicate the feasibility of reliable real-time bathymetry imaging during a single flight of UAS.
\end{abstract}


\section{Introduction}\label{intro}
Accurately measuring nearshore bathymetry on open-coast sandy barred beaches is important for understanding regional sediment dynamics, sandbar migration, and accurate prediction of surf-zone hydrodynamics and wave-driven coastal flooding events. Along wave-dominated coastlines, sandbars migrate frequently in response to changing wave conditions ~\citep{elgar2001nearshore, plant1999simple}, and measurements of surf-zone bathymetry can rapidly become out of date. Surveying surf-zone bathymetry using traditional vessel-based approaches, however, is particularly challenging along these coasts as breaking waves can make vessel-based surveys dangerous and complicate acoustic data collection in the surf. Remote sensing techniques are an alternative approach that can be a valuable tool for monitoring bathymetry changes in the nearshore zone. Unfortunately, in the surf-zone, breaking waves and increased suspended sediment can limit light penetration, effecting the coverage and/or accuracy of airborne lidar or multi-spectral approaches. For this reason, wave-based bathymetry inversions, which use remote sensors like electro-optical, thermal, or radar imagery to measure either the movement of waves through the surf (wave speed or celerity) or the location of wave breaking, and convert these measurements to estimates of bathymetry, have become popular alternatives ~\citep{holman2013cbathy, plant2008ocean, van2008beach, holman2014parametric, holman2016parametric, bergsma2019radon, bergsma2018video}. One advantage of these video-based observations, is that the required data can be easily collected from small unmanned aircraft systems (sUAS)~\citep{holman2017surf, brodie2019uas}, which could broaden accessibility and drastically reduce the time and effort typically required for a nearshore bathymetry survey due to their mass production and inexpensive costs. 

Several approaches have been developed to process video-based imagery data and estimate the corresponding bathymetry~\citep{stockdon2000estimation,holland2001application,van2008beach}. Among many approaches, the cBathy~\citep{holman2013cbathy,holman2021updates} is one of the most widely used algorithms consisting of three processing steps in which (1) the full-motion video imagery are analyzed to provide point-wise estimates of dominant wave-numbers and frequencies at predefined grids (Phase 1), (2) the obtained wave-number and frequency pairs are combined using a non-linear solver and inverted to local water depths through the linear dispersion relationship \citep{plant2008ocean} (Phase 2) and (3) sequential estimates of depth, typically collected on shore-based persistent cameras, are temporally averaged via a Kalman filter to reduce the processing error and improve the estimation accuracy (Phase 3). The accuracy of the bathymetry estimation is limited by the (1) the ability to accurately measure wave speed (or frequency and wavenumber) from the remote observations, and (2) the validity of the linear dispersion relationship. As a result, there can be large estimation error in the shallow water region at the onset of wave breaking or close to the shoreline, potentially creating biases in the location of the sandbar ~\citep{brodie2018evaluation}.  The cBathy algorithm was originally developed for use at persistent monitoring towers, and as a result, the ability to temporally average using the Kalman filter helped compensate for spatio-temporal variability in the accuracy of phase 2 results due to wave breaking, enabling the ability to quantify seasonal sandbar migrations \citep{brodie2018evaluation, bergsma2019storm}. 

The recent applications of cBathy to sUAS, however, changes the most common sampling scheme, as hourly, continuous data streams are no-longer available to support phase 3 of cBathy. Unfortunately, cBathy phase 2 results can often have patchy spatial coverage and high error \citep{brodie2018evaluation}, motivating the development of robust and reliable batch data fusion-based inversion approaches from a single UAS flight with limited dwell time. In other words, there is a need for new robust data fusion and inversion techniques to improve depth estimates since running average bathymetry estimation from data collected over the several days is no longer available. 

An alternate method for estimating a region's bathymetric shape is a barred parametric beach approximation \citep{holman2014parametric, holman2016parametric} or Parametric Beach Tool (PBT), a user-driven method that produces bathymetries with similar parametric shapes and feature length scales as historical bathymetric surveys, utilizing ortho-rectified nadir imagery to identify the cross-shore location of the shoreline and sandbar (from wave-breaking) and estimations of local coastal slopes. The PBT requires the user to input regional characteristics of beach and offshore slope, an offshore water depth, and spatially variable coordinates of the shoreline and bar, which are typically derived from time-averaged (referred to as timex) or variance (referred to as var) images, of the surf.   While the shoreline and sandbar location are derived from observations, the inputs and assumptions of the chosen parametric form can greatly effect the final accuracy and must be well-known for a region.  Though simplistic in it's approach, bathymetry generated using the PBT creates a continuous bathymetric surface, typically accurately resolves the cross-shore location of the sandbar(s), and as been shown be viable for nearshore modeling ~\citep{holman2016parametric}.  

For reliable nearshore characterization using optical measurements from a single UAS flight, we test a combination of 1) dispersion relationship-based inversion from video-derived celerity sub-hourly observations and 2) var image-based parametric bathymetry identification in order to develop a robust bathymetry estimation approach. This approach uses the celerity-derived depth estimates when their accuracy is high, and uses the bathymetric shape information from the PBT when confidence is low. Our goal is to address some of the limitations in each of these approaches by developing an improved bathymetry estimation method in a statistical framework that can optimally ``blend'' or fuse the information from wave celerity and var image-derived bathymetry profiles. 

Since these data will likely be collected by UAS, we also consider a fast and scalable algorithm that enables near real-time bathymetry identification. In many applications such as military operations and emergency responses in nearshore areas, rapid and accurate estimation of the nearshore environment is necessary. In these cases, operations can be performed following a two-stage procedure: 1) optical measurement collection over a site and 2) subsequent computation of near real-time bathymetry estimation on the flight back. With its ``online'' method for on-the-fly wave celerity extraction \citep[e.g.,][]{gawehn2021self} and parameterized bathymetry construction, the computationally fast and storage-efficient bathymetry estimation proposed in this paper can in principle be deployed on a UAS for a fast turnout of the bathymetry estimation at arrival. This configuration would also allow field practitioners to design a follow-on survey flight that would minimize the estimation error based on the error estimate and/or estimation uncertainty.  

The next section describes the brief explanation of PBT and cBathy and the details of the proposed algorithm, which we refer to as ``blending'' due to its combination of PBT and cBathy Phase 1 celerity data. In Section \ref{sec:applications}, we evaluate the accuracy and sensitivity of our proposed framework using Argus tower images taken at the US Army Corps of Engineers Field Research Facility (USACE-FRF) in Duck, North Carolina, USA. Specifically, we compare our results to ground-truth surveys taken on the same dates as the images for three data collection dates in 2017 when ground-truth surveys were performed. Discussion of the results and conclusions based on our work follow in Sections \ref{sec:disc} and \ref{sec:conc}, respectively. 

\section{Methods}
\subsection{Parametric Beach Tool}
\label{sec:pbt}

The PBT is a user-driven method that uses ortho-rectified nadir imagery and estimates of local coastal slopes to approximate a region's bathymetric shape using parametric features and length scales that are similar to historical bathymetric surveys. This background shape profile is then augmented with a sandbar and positioned relative to the current shoreline, based on observations of wave breaking in imagery. 

The PBT algorithm is based on a general equation described in~\citet{holman2014parametric}. It combines a background parametric shape with a bar at a specified location. 
The general equation in 1D, $h$, can be described as:
\begin{align}
    h(x,t) &= h_0(x) + h_{bar}(h_0, t)\nonumber \\
    h_0(x) &= \alpha + \beta x + \gamma \exp{(-kx)}\nonumber \\
    h_{bar}&=-S(h_0)R(t)\cos[\theta(h_0)-\psi(t)]
    \label{eq:pbt}
\end{align}
where $h_0$ describes the background parametric shape and $h_{bar}$ describes the bar shape.


The unknown background shape constants $\alpha$, $\beta$, $k$, and $\gamma$ for $h_0$ are empirically solved using boundary conditions. Assuming a shore-based coordinate system (the model grid is shifted in the PBT algorithm so that $h = 0$ at $x = 0$), $h_{0}$ can be re-written as:
\begin{align}
\label{eq:h_0-v1}
    h_0(x) = \gamma[\exp{(-kx)}-1] + \beta x
\end{align}
where $\beta$ can be assumed to be the offshore beach slope ($\beta_0$). Then, an assumed depth ($h^{\prime}$) at a known offshore location ($x^{\prime}$) can be used and equation \eqref{eq:h_0-v1} can be re-written as
\begin{align}
\label{eq:h_0-offshore}
    h^{\prime} = \gamma [\exp{(-kx^{\prime})}-1] + \beta_{0}x^{\prime}
\end{align} 
This point $(x^{\prime}, h^{\prime})$ should be representative of the background profile and thus a point offshore of the active breaking zone should be used. To finish solving for boundary conditions one can take the derivative of equation \eqref{eq:h_0-v1} where
\begin{align}
    \frac{dh_0}{dx} = -\gamma k \exp{-kx} + \beta_0
\end{align} 
and at the shoreline ($x=0$) setting $\frac{dh_0}{dx}$ to shoreline slope $\beta_s$ we can use
\begin{align}
    \beta_s = -\gamma k  + \beta_0
\end{align}
combined with equation \eqref{eq:h_0-offshore} to solve for $\gamma$ and $k$  numerically. 

Following~\citep{holman2014parametric}, bar amplitude changes were assumed to be small and so $R(t)$ was set to be a constant value of 1. In order to capture the spatial envelope of bar amplitude, the following expression for $S(h_0)$ was used 
\begin{align}
\label{eq:S-def}
S = \delta \frac{x}{x_{off}} + (S_{max} - \delta*\frac{x_{max}}{x_{off}})\exp{\left[\frac{-\left((1-\frac{h_0-h_{shore}}{h_{sea}-h_{shore}})^a-b\right)^2}{c}\right]}
\end{align}
where $\delta$ is the noise floor threshold which is taken empirically as 0.3 \citep{holman2014parametric}. $x_{off}$ is the absolute distance between $x^{\prime}$and the user entered $x_{shore}$ interpolated to the output grid. $S_{max} =0.2h_{sea}$ (where $h_{sea}$ which is the depth of the seaward limit of bar activity), a relationship determined empirically \citep{holman2014parametric}. $x_{max}$ is the cross-shore location where the exponential is at its max. $h_{shore}$ is a site specific value that is the landward extent of bar activity. $a=0.53$, $b = 0.57$, and $c = 0.09$ and are found empirically \citep{holman2014parametric}. By including the factor of $\frac{x}{x_{off}}$ multiplied by $\delta$ as a linear ramp it ensures that there is smoothness at the boundaries of the bar when combined with the background parametric shape.

The phase structure for the cosine bar form is defined as
\begin{align}
\label{eq:theta-def}
\theta=\int_{x_{\mbox{\tiny off}}}^{x}\frac{2\pi}{L(x)}dx
\end{align}
where integration proceeds from the offshore limit of the domain, $x_{\mbox{\tiny off}}$ to each point $x$, and $L(x)$ is given implicitly as~\citep{holman2014parametric}
\begin{align}
\label{eq:L-def}
L(h_0(x))= 100\exp{0.27h_0(x)}
\end{align}
Here, the constants $100$ and $0.27$ are based on a best-fit analysis from \citep{ruessink2003}. The PBT expression for the bar's temporal phase is defined by noting that with a cosine form, the bar crest occurs when the argument of the cosine expression is zero. In other words, assuming that a bar position, $x_b$ can be identified, $\psi(t)$ can be found by requiring
\begin{align}
\psi(t)=\theta(x_b)
\end{align}
where $\theta(x_b)$ is determined from equation \eqref{eq:theta-def} and \eqref{eq:L-def} evaluated at $x_b$. 

General and site specific parameters are summarized in table \ref{tab:param_pbt}.

For the nearshore sites considered below, the alongshore variability is significant and so the 2D extension of the PBT is required~\citep{holman2016parametric}. In this approach, using a subset of shoreline points the line of best fit is found, from which a normal line is taken from a point $(x_p, y_p)$ to the line of best fit. The distance $d$ is found from this point $(x_p, y_p)$ to where the normal line intersects the shoreline at $(x_0,y_0)$. Along $d$ the 1D barred profile is found as described above and the depth along that line is interpolated along $d$. This process is then repeated over the entire grid to perform the 2D bathymetry identification from selected bar and shoreline points. Additional details on the 2D algorithm can be found in~\citet{holman2016parametric}.

\subsection{Bathymetry Estimation}
We next describe an approach to blend an initial-guess bathymetry (from, for example, the PBT) with wave celerity-based inversion.

To infer water depths from wave celerity data (i.e., video-derived frequency-dependent wave numbers), we consider the following observation equation: 
\begin{equation}
\mathbf{k}_{f} = \mathbf{f}(\mathbf{h}) 
\label{eq:obs_eqn}
\end{equation}
where $\mathbf{k}_f$ is the $n_{obs}$ wave number observation vector at frequency $f$, $\mathbf{h}$ is the $n_{bathy}$ unknown water depth vector, and the observation function $\mathbf{f}$ can be implicitly solved through the dispersion equation based on linear wave theory:
\begin{equation}
    \left(2 \pi f_j\right)^2 = g k_{f_j} \tanh{\left(k_{f_j} h_i\right)}
    \label{eq:disp}
\end{equation}
where $h_i$ and $k_{f_j}$ are the local water depth [$m$] and the corresponding wave number  [$\textrm{m}^{-1}$] at the frequency $f_j$ at the location $i$, respectively. 

The inverse problem in Eq.~\ref{eq:disp} can be formulated within a statistical framework. With the initial bathymetry guess, for example, from the PBT, Bayes' rule allows us to evaluate a probability distribution (i.e., posterior distribution) of $\mathbf{h}$ by incorporating the celerity data and: 
\begin{equation}
p(\mathbf{h} | \mathbf{k}_f) \propto p(\mathbf{k}_f | \mathbf{h}) p'(\mathbf{h}) 
\label{eq:bayes_rule}
\end{equation}
where $p'(\mathbf{h})$ represent the prior probability of the bathymetry, and $p(\mathbf{k}_f|\mathbf{h})$ is the likelihood. Here, we assume a prior probability, $p(\mathbf{h}_{prior})$ that follows a Gaussian distribution with mean $\mathbf{h}_{PBT}$ from the PBT and a prior covariance $\mathbf{{C}_{hh}}$, which can be assigned, for example, through variogram analysis with historical data.  

The posterior pdf can be constructed through Eqs.~\ref{eq:obs_eqn}--\ref{eq:bayes_rule} and its negative log-posterior $-\ln p(\mathbf{h}|\mathbf{k}_f)$ is used in this study as an objective function $L(\mathbf{h})$ to determine the bathymetry estimate and its associated uncertainty:
\begin{equation}
\label{eq:obj_cbathy}
    -\ln p(\mathbf{h}|\mathbf{k}_f) = \frac{1}{2}\left(\bm{k}_f - \mathbf{f}(\mathbf{h})\right)^{\top} \mathbf{C}_{\mathbf{kk}}^{-1} \left(\mathbf{k}_f - \mathbf{f}(\mathbf{h})\right) + \frac{1}{2}\left( \mathbf{h} -  \mathbf{h}_{prior} \right)^{\top} \mathbf{C}_{\mathbf{hh}}^{-1} \left( \mathbf{h} - \mathbf{h}_{prior} \right) 
\end{equation}
where $\mathbf{C}_{\mathbf{hh}}$ is the prior covariance matrix and $\mathbf{C}_{\mathbf{kk}}$ is the observation error matrix which is typically defined as a diagonal matrix with elements given as the variance of the measurement error $\sigma_{k_{f_j,i}}^2$ at the frequency $f_j$ and the location $i$. We will explain how to determine $\sigma_{k_{f_j,i}}^2$ later in this subsection. By minimizing the negative loglikelihood function $-\ln p(\mathbf{h}|\mathbf{k}_f)$ with respect to $\mathbf{h}$, we obtain the maximum a posterior (MAP) or most likely value $\hat{\mathbf{h}}$ with the estimation uncertainty through iterative linearization. 

Unfortunately, operations associated with dense covariance matrices such as $\mathbf{{C}_{kk}}$ and are typically computationally prohibitive and so make real-time assimilation infeasible. In addition, since we collect wave number observations in multiple frequencies, it is expected that the number of observations will be larger than the number of unknowns. This leads to a least-squares type problem instead of standard inverse modeling. For computational efficiency, we will then use a Gauss-Newton approach that solves an $n_{bathy} \times n_{bathy}$ linear system iteratively instead of more widely used Kalman-type approaches which lead to an $n_{obs} \times n_{obs}$ system~\citep{evensen2007}. Specifically, in each iteration we update the bathymetry estimate until convergence as:
\begin{equation}
\mathbf{h}^{k+1} = \mathbf{h}^{k} + \Delta \mathbf{h}^{k} 
\label{eq:GN_update}
\end{equation}
and the update $\Delta \mathbf{h}^{k}$ at the $k$-th iteration is determined from 
\begin{equation}
\label{eq:Gauss-Newton}
\mathbf{H}_{GN}^{k} \Delta \mathbf{h}^{k} = -\mathbf{g}^{k} 
\end{equation}
where $\mathbf{H}_{GN}$ is the Hessian matrix with Gauss-Newton (GN) approximation and $\mathbf{g}$ is the gradient:
$$
\mathbf{g} = \frac{\partial L}{\partial \mathbf{h}},\;\;\;\mathbf{H}_{ij} = \frac{\partial^2 L}{\partial \mathbf{h}_i \partial \mathbf{h}_j} \approx \mathbf{H}_{GN}^{k} = \mathbf{J}^{T}\mathbf{C_{kk}}^{-1}\mathbf{J} + \mathbf{{C}_{hh}^{-1}}
$$
Here, $\mathbf{J}$ is the Jacobian of the forward model $\mathbf{f}$, i.e., the (local) model sensitivity of wave observations to the depth:
\begin{equation}
    \mathbf{J}_{ij} = \frac{\partial \mathbf{f}_i}{\partial \mathbf{h}_i} = \frac{\partial k_i}{\partial h_j} = \frac{-k_i^2 \sech^2(k_i h_j)}{\tanh(k_i h_j) + k_i h_j \sech^2(k_i h_j)}
\end{equation}
Note that the Jacobian $\mathbf{J}$ is a sparse matrix due to the point-wise linear dispersion relationship as in~\ref{eq:disp}, thus in the implementation, the Hessian matrix can be computed efficiently without fully constructing the Jacobian matrix $\mathbf{J}$. 

Since the solution to the $n_{bathy} \times n_{bathy}$ linear system in Eq.~\ref{eq:Gauss-Newton} becomes computationally expensive for large-area survey-based bathymetry identification on the fine grid, it is desirable to solve Eq.~\ref{eq:Gauss-Newton} without direct $\mathbf{H}_{GN}$ construction and computation of Eq.~\ref{eq:Gauss-Newton}. One way to achieve this is to use iterative methods, e.g., Krylov subspace methods. Here, we use the conjugate gradient (CG) method because of the symmetric Hessian matrix. We also accelerate the inversion further by replacing the dense inverse covariance matrix $\mathbf{C_{hh}^{-1}}$ with a sparse Laplacian matrix, i.e,  $\mathbf{\mathbf{C_{hh}^{-1}}} = \mathbf{L}\mathbf{L^{T}}$, where
\begin{equation}
\mathbf{L} =  \begin{bmatrix}
1 & -1 &  &  & \\
 & 1 & -1 &  & \\
 &  & \ddots & \ddots & \\
 &  &  & 1 & -1 \\
\end{bmatrix}
\label{eq:fd}
\end{equation}

Then, we can use an (inner) CG - (outer) GN method, where CG is applied to solve Eq.~\ref{eq:Gauss-Newton} and update the GN solutions in Eq.~\ref{eq:GN_update}. This CG-GN method requires a small memory footprint and can be suitable for near real-time applications when the Hessian-vector product can be efficiently computed ~\citep{lee2016scalable,haber2001preconditioned}. This is the case when we utilize the sparse matrices $\mathbf{J}$ and $\mathbf{L}$ whose non-zero elements are $\mathcal{O}\left(n_{obs}\right)$, meaning that the number of non-zero elements increases linearly with the number of observations (for our case $n_{obs} > n_{bathy}$), and thus associated arithmetic computation cost is $\mathcal{O}\left(n_{obs}\right)$. 

The Laplacian matrix is also called as the first-order Tikhonov regularization matrix \citep{tikhonov1977solutions}. In two-dimensional applications, the matrix  is equivalent to that created from a generalized covariance function~\citep{kitanidis1999generalized} with log-linear covariance model:
\begin{equation}
\mathbf{C}_{{\mathbf{h}_i}{\mathbf{h}_j}} = \text{Cov}(\mathbf{h}_i,\mathbf{h}_j) = - \ln |\mathbf{h}_i-\mathbf{h}_j|
\end{equation}
where $\mathbf{h}_i, \mathbf{h}_j$ are the water depth at the location i and j, respectively. thus the log-linear covariance model is chosen for our inversion. Eq.~\ref{eq:Gauss-Newton} becomes:
\begin{equation}
\mathbf{H}_{GN}^{k} \Delta \mathbf{h}^{k} = \left(\mathbf{J}^{T}\mathbf{C}^{-1}_{\bm{kk}}\mathbf{J} + \mathbf{C_{hh}^{-1}}\right) \Delta \mathbf{h}^{k} = \left(\mathbf{J}^{\top}\mathbf{R}^{-1}\mathbf{J} + \mathbf{L}^{\top}\mathbf{L}\right) \Delta \mathbf{h}^{k} = -\mathbf{g}^{k} 
\end{equation}
which can be solved efficiently. The GN method at $k$-th iteration yields the update as
\begin{equation}
    \mathbf{h}^{k} =  \mathbf{h}^{k-1} + \alpha^{k-1} \left(\mathbf{J}^{T} \mathbf{R}^{-1} \mathbf{J} + \mathbf{L}\mathbf{L}^{T}\right)^{-1} \mathbf{J}^{T} \mathbf{R}^{-1}\left(\mathbf{k}_f - \mathbf{f}\left(\mathbf{h}^{k-1}\right) - \mathbf{L}\mathbf{L}^{T}\left(\mathbf{h} - \mathbf{h}_{\text{PBT}}\right)\right)
    \label{eq:update}
\end{equation}
The proposed method is implemented in Python (see Appendix \ref{sec:appendixB}) and its accuracy and scalability will be tested and discussed in the following sections.

\subsection{Observation screening and error calibration}
Accurate assessment of observation error plays a key role role in the blending process. Here, we rely on cBathy error statistics both to screen outliers of wave number observations and to set the error variance $\sigma^2$ used to determine $\mathbf{C}_{\mathbf{kk}}$ in Eq.~\ref{eq:obj_cbathy}. The available statistics from cBathy Phase 1 are \texttt{skill} of the fit $s$, the first dominant eigenvalue of the covariance \texttt{lambda}, \texttt{kErr}, the 95\% confidence interval of the estimated frequency dependent $k$, and \texttt{hTempErr}, the 95\% confidence interval of the estimated frequency-dependent water depth. \texttt{skill} and \texttt{lambda} are used to determine the quality of the wave celerity data following the cBathy algorithm \citep{holman2013cbathy}, and we discard observations where \texttt{skill} $<$ 0.5 and \texttt{lambda} $<$ 10 following the default cBathy Phase 1 criteria. We then use \texttt{kErr} to determine the spatially distributed error in frequency-dependent $k_f$ observations. Specifically, we use the (prior) error variance $\sigma^2 = \left(\alpha \cdot kErr\right)^2$ where $\alpha$ is a regularization parameter to be calibrated. This leads to the observation error matrix given by Eq.~\ref{eq:C_kk}. There are a number of ways to calibrate $\alpha$, including methods such as the L-curve method or cross-validation ~\citep{hansen1992analysis,wahba1990spline}. From our initial calibration tests, $3 \le \alpha \le 8$ gave similar satisfactory results. We set $\alpha=5$ in our applications presented below. 
\begin{equation}
\mathbf{C}_{\mathbf{kk}} =  \begin{bmatrix}
\sigma_{1,1}^2 &  &  &  & \\
 & \sigma_{1,2}^2 &  &  & \\
 &  & \ddots & & \\
 &  & & \sigma_{n_{obs},3}^2 & \\
 &  &  &  & \sigma_{n_{obs},4}^2 \\
\end{bmatrix}
=  \alpha^2 \begin{bmatrix}
\textrm{kErr}_{1,1}^2 &  &  &  & \\
 & \textrm{kErr}_{1,1}^2 &  &  & \\
 &  & \ddots & & \\
 &  & & \textrm{kErr}_{n_{obs},3}^2 & \\
 &  &  &  & \textrm{kErr}_{n_{obs},4}^2 \\
\end{bmatrix}
\label{eq:C_kk}
\end{equation}

\section{Applications}
\label{sec:applications}
We test our proposed framework with observations taken from the U.S. Army Corps of Engineer Field Research Facility (FRF) in Duck, North Carolina. The FRF site features a relatively straight coastline and is considered representative of many U.S. beaches in terms of wave climate, storm exposure, and sand size. The FRF site has supported numerous field experiments for understanding wave dynamics and coastal processes~\citep{birkemeier1996delilah,potter2022},and resulting data sets including extensive survey and continuously monitored wave and tide data have been collected and maintained by US Army Engineering Research and Development Center Coastal and Hydraulic Laboratory (\url{http://www.frf.usace.army.mil}). The FRF beach has a tidal range of 0.5–2.0 m and video images of the nearshore region of the FRF site has been collected with an Argus coastal monitoring station located 43 m above mean sea level on a tower. This imagery was used for both parts of the proposed blending framework. Time-series imagery data were collected at a 2 Hz sample rate for 17 min every half hour to estimate frequency and wave number at a spatial resolution of 5 and 10 m in the cross- and alongshore directions, respectively, over the analysis domain, which measured 800 m in the cross-shore an 1500 m in alongshore direction. The time series imagery can be processed to create different products including timex and var images. 10 minute time-averages of pixel intensity at the start of each half-hour were used to construct timex images. The variance is also calculated at each pixel to construct var images which show areas of high change. 


\begin{figure}
    \centering
    \includegraphics[width=0.5\textwidth]{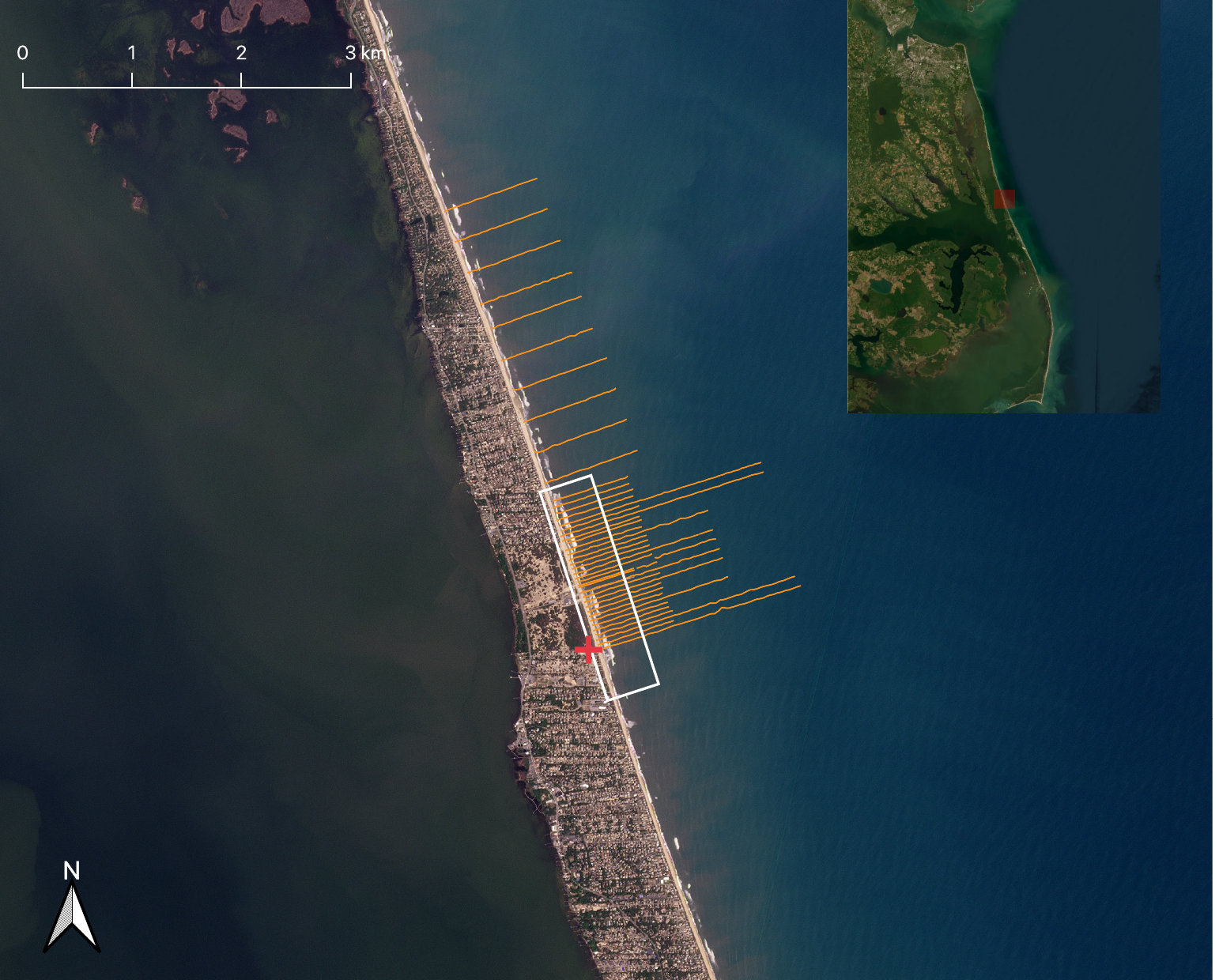}
    \caption{Site map with the extent of Argus imagery outlined in the white box, the red X is the origin in FRF coordinates, and orange lines are monthly survey transect lines. In the top right corner the inset shows a zoomed out picture of the Outer Banks in NC with the site in the red box.}
    \label{fig:sitemapzoomed}
\end{figure}
We chose the imagery data sets taken on three dates, March 27 17:00, 2017, November 22 20:00, 2017, and October 6 16:30, 2019 based on the availability of in-situ survey data. Table \ref{tab:weather} shows wave height data and peak period from the US Army CHL Data Server \citep{waves} and weather conditions from National Weather Service \citep{weather} on the test dates. These selected dates cover various ranges of wave height, wind speed and peak period conditions that occur at the FRF. The in-situ surveys were conducted with a Lighter Amphibious Resupply Cargo (LARC) vessel and the Coastal Research Amphibious Buggy (CRAB), and the measured seafloor elevations (relative to NAVD88) were interpolated to match the cBathy grid points \citep{forte2017nearshore}. Note that at Duck, in-situ surveys were not performed with waves larger than 2.0 m because of the logistical challenges of operating in the presence of large waves \citep{brodie2018evaluation}. 
\begin{table}[!ht]
    \centering
    \begin{tabular}{|c|c|c|c|c|c|}
        \hline
         date & time  & wind [gusts] & direction & significant wave height   & peak period \\
              & (GMT) &     (mph)    &           &           (m)             &     (s)     \\
         \hline
         3/27/2017 & 17:00 & 6 [9] & 232$^{\circ}$SW & 0.87 & 7.71  \\
         \hline
         11/22/2017 & 20:00 & 19 [23] & 7$^{\circ}$N & 1.89 & 6.29  \\
         \hline
         10/6/2019 & 16:30 & 18 [20] & 133$^{\circ}$SE & 1.25 &  9.93 \\
         \hline
    \end{tabular}
    \caption{Wave conditions from US Army CHL Data Server \citep{waves} and weather conditions from National Weather Service \citep{weather} on the test dates}
    \label{tab:weather}
\end{table}
We evaluated the accuracy and sensitivity of the PBT timex and var images by comparing it to ground-truth surveys taken on the same dates as the images. The PBT requires the user to input spatially variable coordinates of the shoreline and bar. The selection of the shoreline and bar location of timex and var images was compared with the features of in-situ surveys to determine the optimal characteristics in images that relate to these parameters. Points were selected at varied cross-shore locations along the bar break (inner edge, middle, outer edge) for comparison. The different image types and selected cross-shore location of the bar break were compared at all survey transects at the FRF (not shown). Among all five dates, the bar was most accurately located when following the outermost edge of the breaking with similar accuracy for the timex and var images, similar to the results of \cite{holman2016parametric}. In the following results, points selected on var images were used, since var images tended to define the offshore boundary of breaking more distinctively than the Timex image, making it easier for manual digitization.

In this application, the PBT bathymetry is set as the prior in the blending framework and also serves as an initial guess of the optimization task. To make sure the blending estimation is robust to the initial bathymetry selection during the iterative optimization in Eq. \ref{eq:GN_update}
, we also tested with different initial bathymetry guesses such as a linearly changing profile with respect to the offshore distance or constant flat profile and confirmed the convergence to a unique final solution regardless of the initial bathymetry choice. Frequency-wavenumbers pairs extracted from cBathy Phase 1~\citep{holman2014parametric,holman2021updates} are used as observations to update and correct the PBT bathmetry into the final estimate consistent with the dispersion equation \eqref{eq:disp}. cBathy-driven estimates (i.e., Phase 2) are not required but may be used for validation purpose. The results are compared to direct surveys and additionally with cBathy phase 2 estimates in the next section~\ref{sec:results}.


Finally, the PBT relies on certain site specific parameters as shown in Equation \ref{eq:pbt}, which have been tuned to the characteristics of the FRF. The offshore slope is defined as 0.0088 and the shoreline slope is 0.10 based on historical data at the FRF \citep{holman2016parametric}. An offshore depth is defined at 700 m alongshore (xFRF = 700 m in the local coordinates of the Duck location) as 7.5 meters, also based on historical data at Duck, which anchors the results to ones that are realistic to Duck historical bathymetry. PBT Parameters used in the application are listed in Table~\ref{tab:param_pbt}.
\begin{table}[htbp!]
    \centering
    
    \begin{tabular}{|c|c|c|}
        \hline
        Parameter & Value & Description \\
        \hline\hline
        \multicolumn{3}{|c|}{General parameters from \citep{holman2014parametric}}\\
        \hline
        $\alpha$, $\gamma$, $k$ & found iteratively & background shape parameters\\
        \hline
        R(t) & 1 & sandbar-wise temporal variability\\
        \hline
        $\delta$ & 0.3 & noise floor threshold\\
        \hline
        a, b, c & 0.53, 0.57, 0.09 & constants \\
        \hline\hline
        \multicolumn{3}{|c|}{Site specific values (values used for Duck, NC \citep{holman2016parametric})}\\
        \hline
        $\beta_{0}$ & 0.088 & asymptotic offshore beach slope\\
        \hline
        $\beta_{s}$ & 0.1 & estimated climatological shoreline slope\\
        \hline
        $h^{\prime}$ & 7.5 m & known deep water value\\
        \hline 
        $x^{\prime}$ & 700 m & cross-shore distance of known deep water value\\
        \hline
        $h_{shore}$ & 0 m & depth of landward extent of bar activity \\
        \hline
        $h_{sea}$ & 4.5 m & depth of seaward extent of bar activity \\
        \hline 
        $x_{off}$ & $|x^{\prime} - x_{shore}|$ & distance between location of seaward extent of bar and shore\\
        \hline\hline
        \multicolumn{3}{|c|}{User entered values}\\
        \hline
        $(x_{shore}, y_{shore})$ & shown on figure \ref{fig:pbt_selection} & coordinates of shoreline (can be one or many)\\
        \hline
        $(x_{bar}, y_{bar})$ & shown on figure \ref{fig:pbt_selection} & coordinates of bar (can be one or many) \\
        \hline
    \end{tabular}
    \caption{Parameters used in the PBT application}
    \label{tab:param_pbt}
\end{table}    

\subsection{PBT-based background bathymetry construction} 
\begin{figure}[htbp!]
    \centering
    \includegraphics[width=\textwidth]{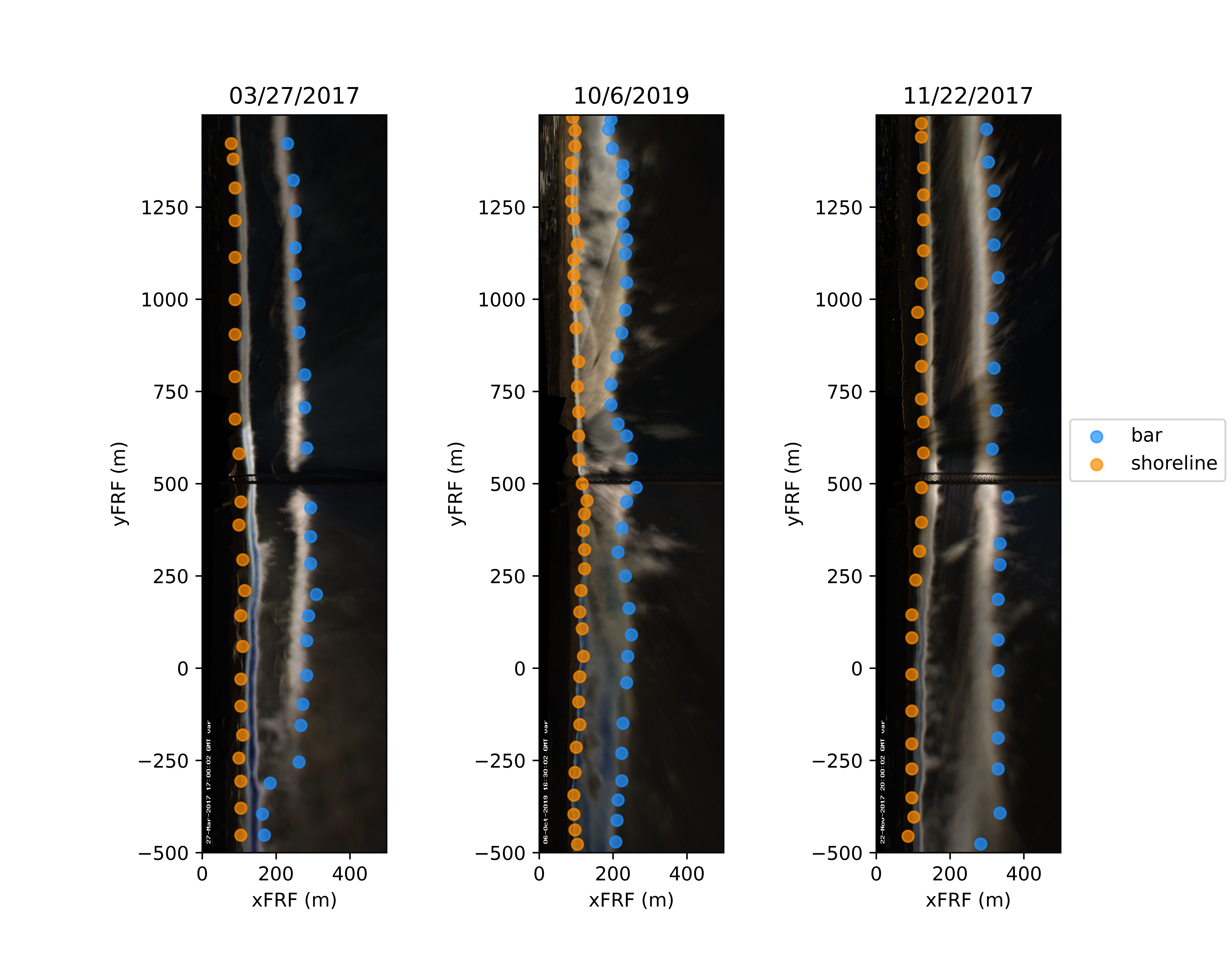}
    \caption{Points used in PBT algorithm that were selected on the var image taken at the FRF. These points were selected using the outer edge of the offshore white water for the barline. These points were then used in the PBT algorithm to get the bathymetric shape. Coordinates are in local FRF space.}
    \label{fig:pbt_selection}
\end{figure}

The PBT was performed following the procedure described in Section \ref{sec:pbt}. In Figure \ref{fig:pbt_selection}, the selected shoreline and bar locations and the Var image they were chosen on is shown. The bar and shoreline were entered into the PBT and the resultant bathymetry was interpolated to the coordinates of the cBathy results for comparison. Once bar and shore coordinates were selected, the PBT takes less than a minute to run in the resolution used in this application.

\subsection{Bathymetry Applications in FRF, Duck NC}
With the PBT-derived bathymetry as an initial guess/prior, the proposed framework incorporates available celerity measurements from imagery to produce the final bathymetry estimate. The computational domain is a 2000-$\textrm{m}$ alongshore and 800-$\textrm{m}$ across-shore area with the xFRX coordinate between 80 and 800 $\textrm{m}$ and the yFRF coordinate between -500 and 1500 $\textrm{m}$. The domain is descritized into 73 x 81 grids with a uniform spacing of $\Delta x$ = 10 $\textrm{m}$ and $\Delta y$ = 25 $\textrm{m}$.

To avoid noisy observations and outliers, we screen cBathy Phase 1 wave celerity points with several criteria. First, the dispersion relationship is used to invert each frequency-wavenumber pair for the water depth and only the observation locations whose inverted water depth is between 0.25 m and 10 m are used in the blending. Additionally, cBathy Phase 1 error statistics~\citep{holman2021updates} are used to choose the observations at locations whose \texttt{skill} is larger than 0.5 and \texttt{lam1} is smaller than 10. 

During the blending, the estimated water depth less than 0.001 $\textrm{m}$ is set to 0.001 $\textrm{m}$ to evaluate the dispersion equation.

Table for number of measurements from image, blending parameters. 
\begin{table}[htbp!]
    \centering
    \begin{tabular}{|c|c|c|}
        \hline
         Parameter & Value & Description  \\
         \hline
          $\Delta x$, $\Delta y$ & 10, 25 $\textrm{m}$ & Pixelwise cross-shore and alongshore spacing \\
         $N_x$, $N_y$, N & 73, 81, 5913 (73 $\times$ 81)  & Total number of unknown bathymetry points  \\
         $h_{min}$, $h_{max}$ & 0.25 $\textrm{m}$, 15 $\textrm{m}$ &  Minimum and maximum acceptable depth\\
         $s_{min}$ & 0.5 & minimum acceptable skill\\
         $\lambda_{min}$ & 10 & minimum acceptable normalized eigvenvalue\\
         $N_{keep}$ & 4 & Number of frequency-wave pairs at a location\\
         \hline
    \end{tabular}
    \label{tab:param_blending}
    \caption{Parameters involved in the applications}
\end{table}

\section{Results}
\label{sec:results}
\subsection{Results from data collected on March 27, 2017 (mild wave height)}
\begin{figure}[htpb!]
    \centering
    \includegraphics[width=\textwidth]{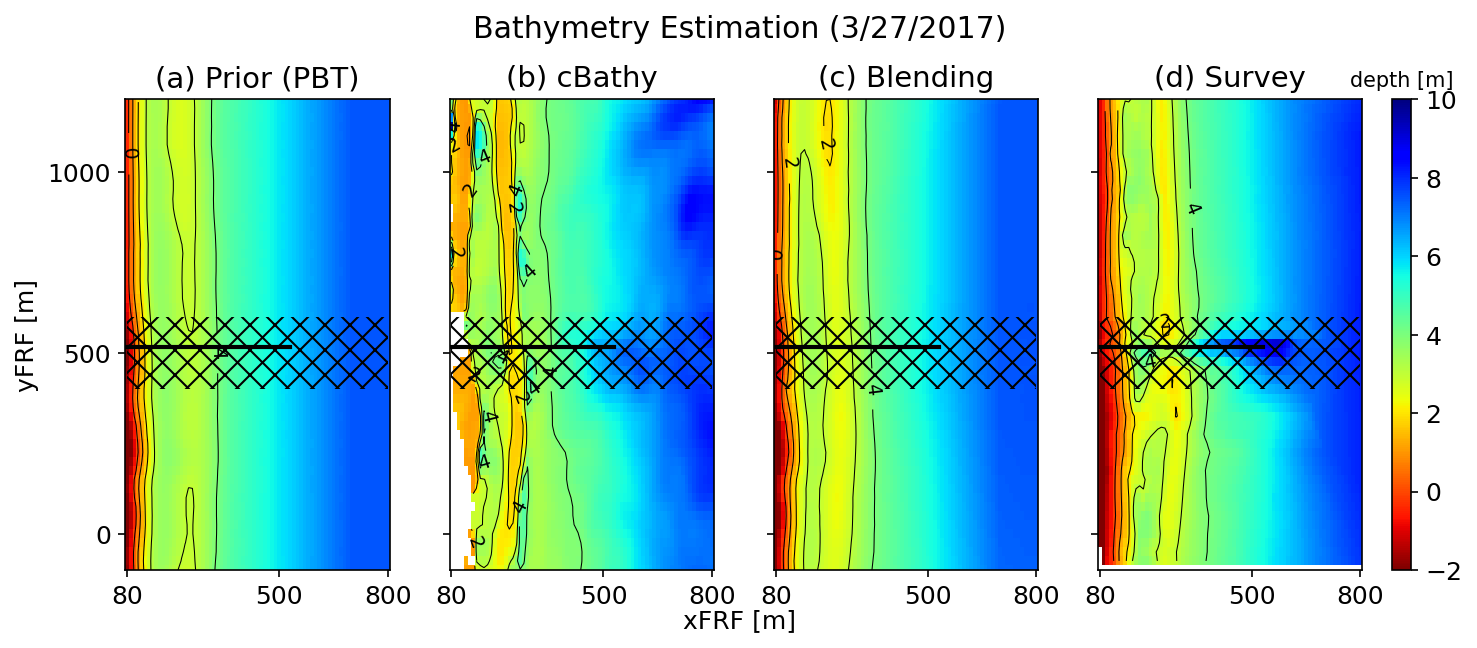}
    \caption{Bathymetry estimates from the data collected on 3/27/2017; black lines represent the FRF research pier (xFRF between 80 $\textrm{m}$ and 530 $\textrm{m}$ and yFRF = 515 $\textrm{m}$) and the pier-affected area is displayed with a hatch pattern (yFRF between 400 and 600 $\textrm{m}$)}
    \label{fig:result_03272017}
\end{figure}

The first case we consider is a bathymetry estimation application on 27 March 2017 17:00, when the weather was relatively calm with slight wave height of 0.87 $\textrm{m}$ as in Table~\ref{tab:weather}. Figure~\ref{fig:result_03272017} (a) shows the prior from PBT, cBathy (Phase 2) estimate  (Figure  ~\ref{fig:result_03272017}b), blending estimate (Figure  ~\ref{fig:result_03272017}c), and survey. The survey data (Figure  ~\ref{fig:result_03272017}d) were projected to the rectangular computation grid using bilinear interpolation and serve as a reference ``true'' bathymetry field in this study. The PBT, which was obtained from calibrated parametric equations of background slope and sandbar from the Var image, yields smooth estimation of bathymetry while the cBathy Phase 2 estimate shows rough spatial variability due to the point-wise dispersion relationship fitting. The proposed blending method takes advantage of the dispersion fitting and var image-based PBT background to produce a better bathymetry estimate with improved accuracy. Specifically, the blending method identifies the trough and sandbar bathymetry between 150 $\textrm{m}$ and 300 $\textrm{m}$ alongshore (150 $\textrm{m}$ $\le$ xFRF $\le$ 300 $\textrm{m}$) well.

\begin{figure}[htpb!]
    \centering
    \includegraphics[width=0.8\textwidth]{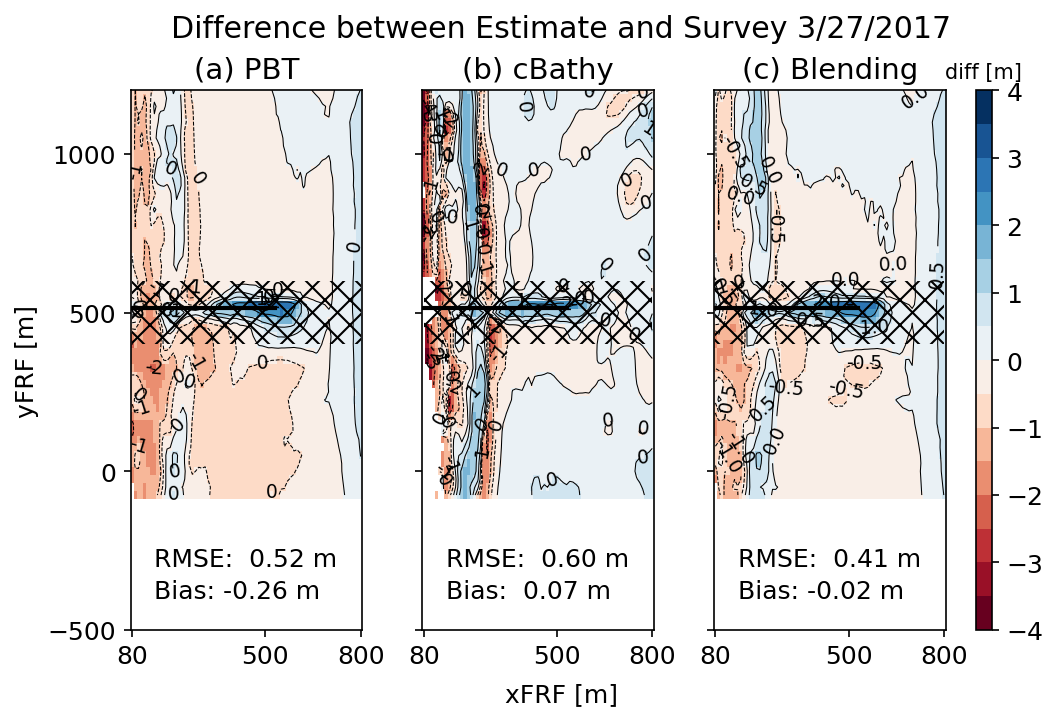}
    \caption{Estimation error for the data set collected on 3/27/2017; the estimation error is computed as the difference between the estimate and surveyed bathymetry over the non-pier area of $400$ $\textrm{m}$ $>$ yFRF or yFRF $> 600$ $\textrm{m}$.}
    \label{fig:diff_03272017}
\end{figure}

Figure~\ref{fig:diff_03272017} shows the estimation errors from the methods considered in this study. The estimation errors are computed as the difference between the surveyed and estimated bathymetry. It is shown that the estimate from the blending method is the most consistent with the surveyed bathymetry and most differences are less then 0.5 $\textrm{m}$. The survey data is used for the estimation error computation in terms of Root Mean Squared Error (RMSE) and bias. The RMSE and bias are computed over the non-pier area of $400$ $\textrm{m}$ $>$ yFRF or yFRF $> 600$ $\textrm{m}$ in order to avoid the known bathymetry anomaly due to the FRF research pier located at yFRF $\sim 500$ m \citep{holman2013cbathy}. These error statistics are also computed only in the underwater area (depth $\ge 0$ $\textrm{m}$) for comparison with the cBathy-derived bathymetry estimate. The identified bathymetry is close to the survey data with RMSE of 0.41 $\textrm{m}$. The RMSE from PBT and cBathy were 0.52 and 0.60 $\textrm{m}$, respectively. The bias of the PBT, cBathy, and blending estimates are -0.26, 0.07 and -0.02 $\textrm{m}$, respectively showing that the bias is improved with our proposed blending approach.

\begin{figure}[htpb!]
    \centering
    \includegraphics[width=\linewidth]{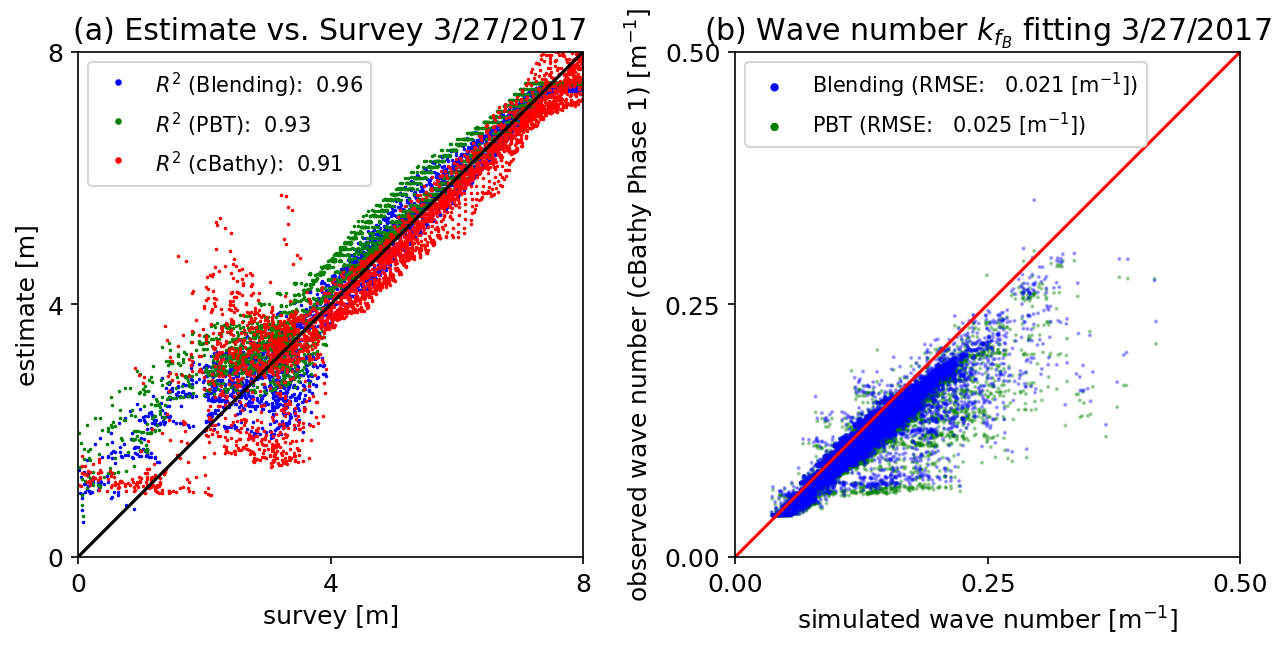}
    \caption{(a) Surveyed vs. estimated water depth, (b) simulated vs. observed wave numbers for PBT and Blending for the 3/27/2017 data set}
    \label{fig:bathy_fit_03272017}
\end{figure}

Figure \ref{fig:bathy_fit_03272017} (a) shows the plot of bathymetry estimation points against survey points, which indicates that the blending method produces the most accurate estimate consistent with the survey over the entire range of the water depth. Correlation coefficients $\textrm{R}^2$ were also computed between surveyed and estimated bathymetry and their values are 0.96, 0.93, and 0.91 for blending, PBT, and cBathy, respectively showing the improvement of blending and PBT. The performance of the cBathy estimate varies with depth \citep[e.g.,][]{holman2013cbathy} and the blending method reduces the bias effectively based on the background information from PBT. The underestimation of the offshore depth in PBT and blending is observed because the offshore boundary depth is set to the in-situ observation of 7.5 $\textrm{m}$ in PBT for its parametric form and the blending method follows the offshore depth from PBT. 
In Figure~\ref{fig:bathy_fit_03272017} (b), the frequency-dependent wave numbers estimated from the blending method and PBT are plotted against the measured wave numbers derived from cBathy Phase 1. The blending method takes the PBT estimate as an initial guess which produced the RMSE of 0.025 $\textrm{m}^{-1}$ and improves the wave number fitting with RMSE of 0.021 $\textrm{m}^{-1}$. 

\begin{figure}[htpb!]
    \centering
    \includegraphics[width=\textwidth]{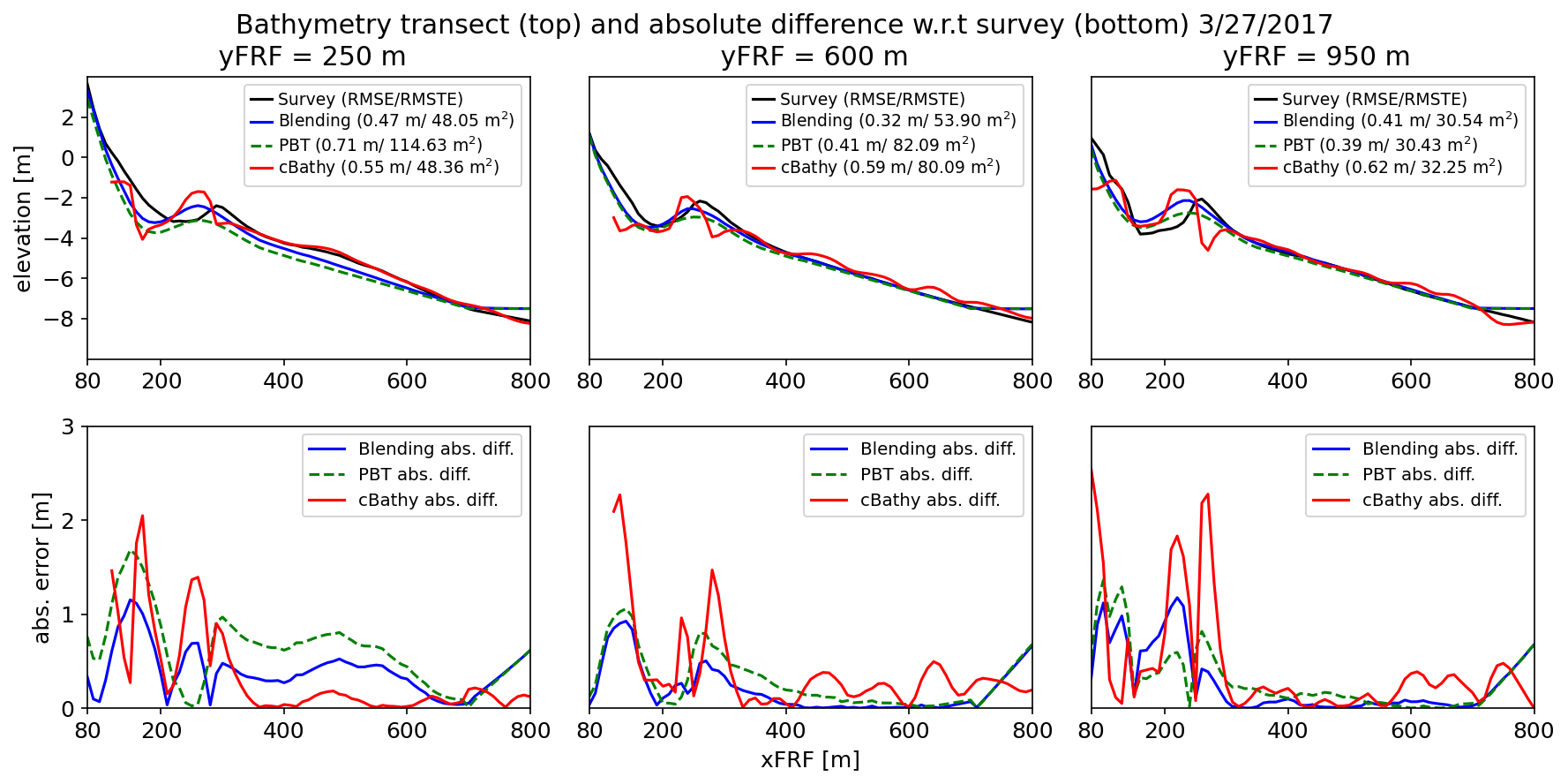}
    \caption{Cross-shore transect estimates (top) and errors (bottom) at yFRF = 250 (left), 600 (middle), and 950 $\textrm{m}$ (right)}
    \label{fig:transect_03272017}
\end{figure}

Figure~\ref{fig:transect_03272017} presents the cross-shore transects in selected locations at yFRF = 250, 600, and 950 $\textrm{m}$ (top) and the corresponding errors (bottom). The errors are computed as the absolute difference between the estimate and survey. Overall, the PBT estimates (blue) generally overestimate the surveyed water depth while the cBathy estimates (red) underestimate the water depth near the sandbar.  The blending estimates closely approximate the survey transects. The cBathy Phase 2's hourly estimates are not constrained to the shoreline location and sandbar details and does not capture underlying morphological features. The bar shapes from the blending estimates look realistic because the blending takes the PBT's parametric form. However, the bar positions from all the methods are biased toward onshore because of the discrepancy between the optical measurements and actual bar locations as discussed in \citet{van2001effect,brodie2018evaluation}. Still, the blending method takes advantage of two different data sources and seems to provide a balanced solution to identify the bar elevations and locations reasonably. 

Morphological features of the sandbar in the survey appear closer to the blending and PBT estimates due to the modeling of the sandbar with the parametric form (Eq. \ref{eq:pbt}) in PBT. cBathy was able to locate the sandbar location, but it does not produce realistic shapes of the bar and trough. The onshore dune area estimated from PBT (elevation of 0 - 4 $\textrm{m}$ above the seawater) also allows the blending method to better identify the shallow inner-surf zone area (up to xFRF = 200 $\textrm{m}$) than that obtained from cBathy, by guiding the slope and its changes around the shallow surf zone.

In Figure~\ref{fig:transect_03272017}, the RMSE values of the estimated transects are reported to show the difference between the reference survey and the estimate. In addition, to qualitatively evaluate morphological similarity between survey and estimated bathymetry especially in the shallow water region, we compute Root Mean Squared Transport Error (RMSTE) proposed by \citet{bosboom2020optimal} for the transect estimates in Figure~\ref{fig:transect_03272017}. RMSTE computes the (quadratic) minimum cost of the sediment transport from one bathymetry to another bathymetry as a proxy of morphological feature comparison instead of point-wise bathymetry comparison as in RMSE. Thus, RMSTE may provide an alternative metric related to how much sands one needs to move from the estimated transect in order to recover the surveyed transect (i.e., earthmover's distance). RMSTE is computed over the shallow water region between the shoreline and the end of sandbar (yFRF $\le 350\; \textrm{m}$). See the Appendix \ref{sec:appendixA} for details. RMSTE of the blending method are much smaller than those of PBT and cBathy at yFRF = 250 $\textrm{m}$ and 600 $\textrm{m}$.

The blending method shows lower RMSE on the three selected transects and the absolute estimation errors from the blending are lowest in general at most locations. In summary, for the data set collected under mild weather and wave conditions, the proposed blending method provides the improved bathymetry estimate and outperforms cBathy and PBT.

\subsection{Results from data collected on October 6, 2019 (moderate wave height/longer peak period)}

\begin{figure}[htpb!]
    \centering
    \includegraphics[width=\textwidth]{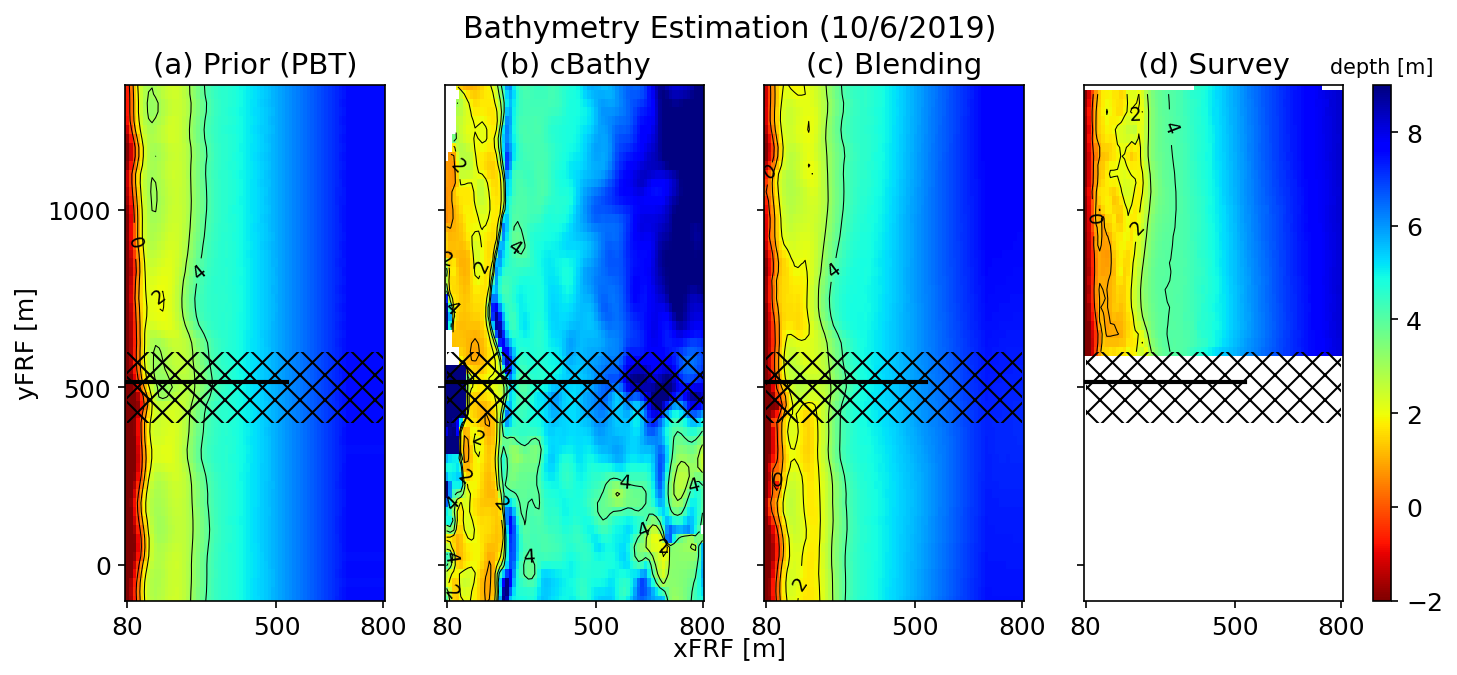}
    \caption{Bathymetry estimates from the data collected on 10/6/2019; black lines represent the FRF research pier (xFRF between 80 m and 530 m and yFRF = 515 m)}
    \label{fig:result_10062019}
\end{figure}

\begin{figure}[htpb!]
    \centering
    \includegraphics[width=0.8\textwidth]{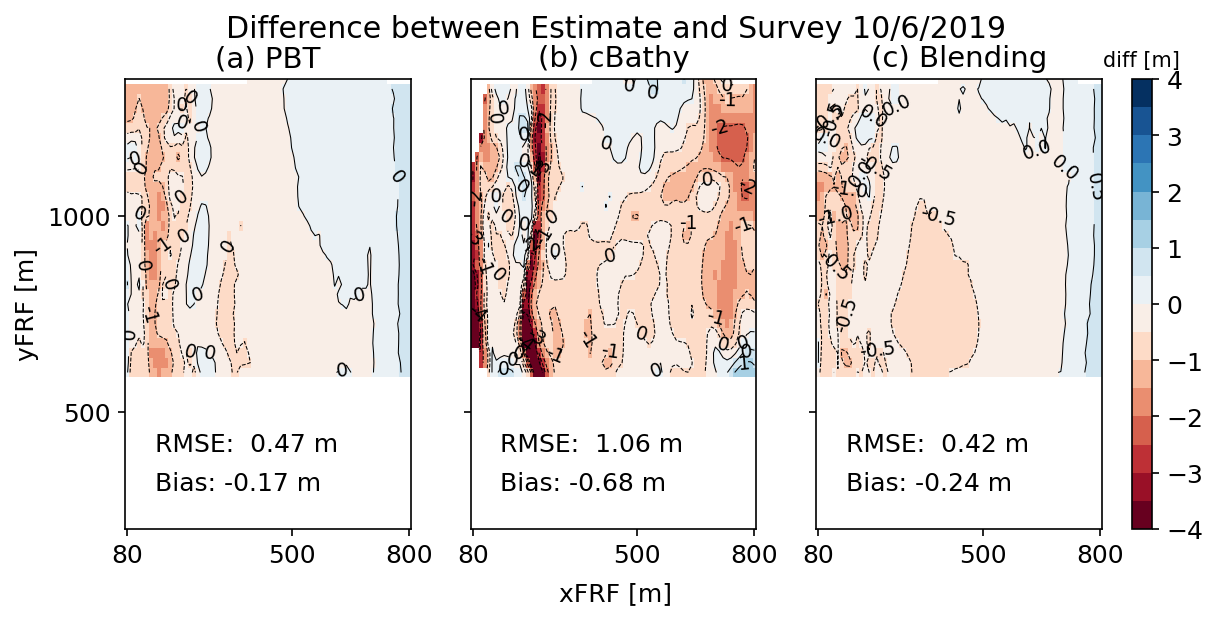}
    \caption{Difference between estimate and survey on 10/6/2019; the estimation error is computed as the difference between the estimate and surveyed bathymetry.}
    \label{fig:diff_10062019}
\end{figure}

The data collected on October 6, 2019 16:30 is considered with a longer peak period of 9.93 $\textrm{s}$ and moderate wave height of 1.25 $\textrm{m}$ as in Table~\ref{tab:weather}. Figure~\ref{fig:result_10062019} shows the bathymetry estimates from PBT, cBathy, the proposed blending method and the survey. The survey data is only available at yFRF between 600 $\textrm{m}$ and 1300 $\textrm{m}$ on this date and the estimation error defined as the difference between the survey and estimated bathymetry is presented in Figure \ref{fig:diff_10062019}. In this case, the PBT results are largely insensitive to the changing conditions, since a good-quality var image is available and provides the selected shoreline and sandbar locations as shown in Figure \ref{fig:pbt_selection} regardless of the weather and wave conditions. Specifically, the PBT method produces a reasonable bathymetry estimate with RMSE of 0.47 $\textrm{m}$, which is comparable to the March 27, 2017 error. However, cBathy-derived bathymetry is significantly poorer with RMSE of 1.06 $\textrm{m}$ and the largest error up to 4 $\textrm{m}$ of the depth over-estimation near the sandbar locations (red in Figure \ref{fig:diff_10062019} (b)). The proposed blending method shows improved results with RMSE of 0.42 $\textrm{m}$ indicating about 10\% RMSE reduction from the PBT (RMSE = 0.47 $\textrm{m}$) while the bias slightly increases by 0.06 $\textrm{m}$.  

\begin{figure}[htpb!]
    \centering
    \includegraphics[width=\textwidth]{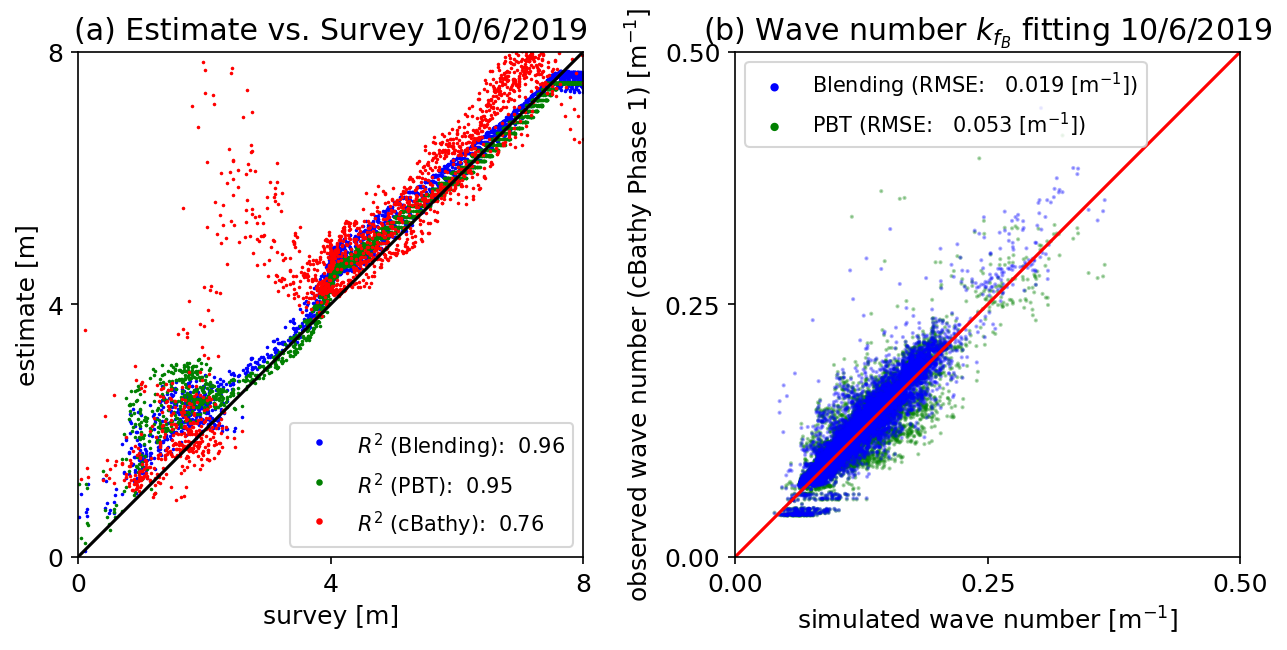}
    \caption{(a) Surveyed vs. estimated water depth, (b) simulated vs. observed wave numbers for PBT and Blending for the 10/6/2019 data set}
    \label{fig:bathy_fit_10062019}
\end{figure}

Figure \ref{fig:bathy_fit_10062019} (a) shows the plot of bathymetry estimation points against survey points. Overall, the PBT (blue) and blending (green) estimates closely predict the survey transects with the coefficient of correlation $R^2$ = 0.95 and 0.96, respectively while cBathy (red) consistently overestimates the depth with $R^2$ of 0.76. It is also observed that over the water depth between 0 $\textrm{m}$ and 3 $\textrm{m}$, estimated points from the blending method (blue dots) are consistently closer than PBT estimates (green dots) to the 45 degree line showing the shallow surf zone (with the water depth up to 3 $\textrm{m}$) is better estimated in the blending method. The blending method incorporates the cBathy-driven wave data to correct the bias and error from PBT in the shallow area resulting in more accurate estimation for shoreline and sandbar. In Figure~\ref{fig:bathy_fit_10062019} (b), the frequency-dependent wave numbers estimated from the blending method and PBT are plotted against the measured wave numbers. PBT produces the RMSE of 0.053 $\textrm{m}^{-1}$ and the blending method improves the wave number fitting with RMSE of 0.019 $\textrm{m}^{-1}$.

\begin{figure}[htpb!]
    \centering
    \includegraphics[width=0.8\textwidth]{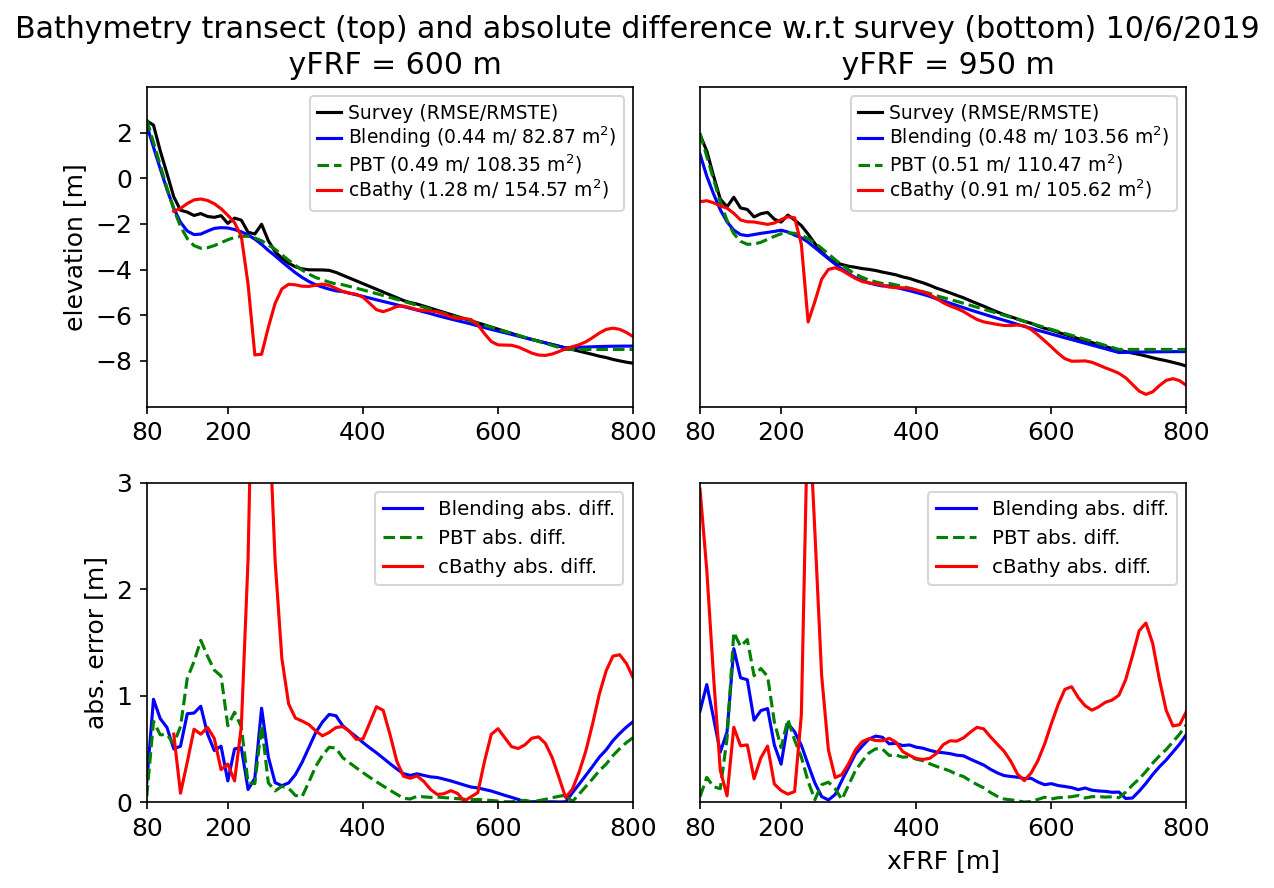}
    \caption{Cross-shore transect estimates (top) and errors (bottom) at yFRF = 600 $\textrm{m}$ (left) and 950 $\textrm{m}$ (right) on 10/6/2019}
    \label{fig:transect_10062019}
\end{figure}

Figure~\ref{fig:transect_10062019} presents the cross-shore transects in selected locations at yFRF = 600 and 950 $\textrm{m}$ (top) and the corresponding errors (bottom) from the imagery data collected on 10/06/2019. Overall, the PBT and cBathy under- and overestimate the depth in most locations of the shallow water region while the blending method closely estimates the survey transects by leveraging the two different data sources. As we observed in the previous Figure \ref{fig:bathy_fit_10062019} (a), the (sandbar crest and trough) area that PBT misses (green dots with the depth between 1 m and 3 m in Figure \ref{fig:bathy_fit_10062019} (a)) are better characterized in the blending estimate because of the use of cBathy celerity data. The blending method shows the lowest RMSE and RMSTE on the selected transects. The performance of the blending method is promising beyond the mild wave height condition --- under the moderate wave height and longer peak period with relatively high wind condition. 


\subsection{Results from data collected on November 22, 2017 (relatively high wave height)}

Lastly, we consider the data collected on November 22, 2017 20:00 in the winter period with moderate wave height of 1.89 $\textrm{m}$ as in Table~\ref{tab:weather}. This wave height is almost a maximum limit at which vessel-based in-situ surveys can be performed, and validating our method with the data higher than 2 $\textrm{m}$ is beyond the scope of our study. 

\begin{figure}[h!]
    \centering
    \includegraphics[width=\textwidth]{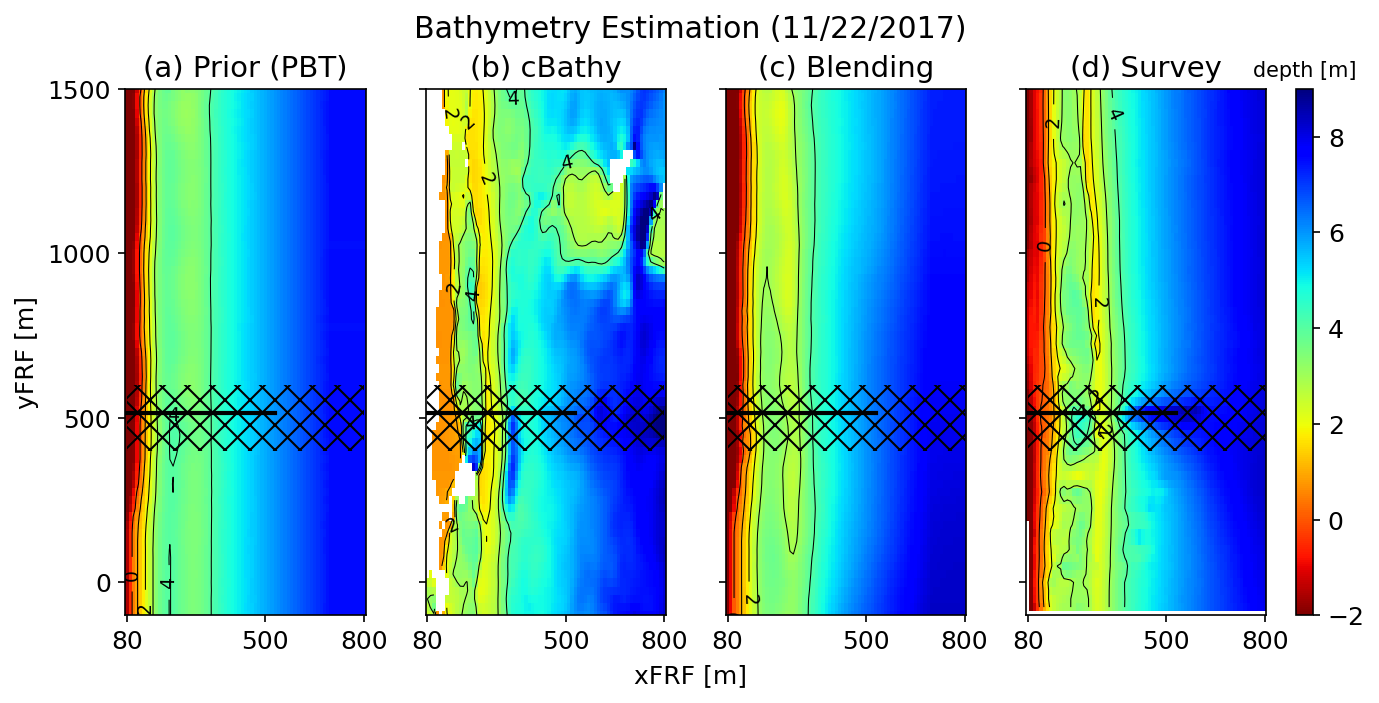}
    \caption{Bathymetry estimates from the data collected on 11/22/2017; black lines represent the FRF research pier (xFRF between 80 m and 530 m and yFRF = 515 m)}
    \label{fig:result_11222017}
\end{figure}

Figure~\ref{fig:result_11222017} shows the prior from PBT, cBathy Phase 2 estimate, blending estimate, and survey. The PBT estimate is reliably obtained based on shoreline and sandbar locations from the Var image in Figure \ref{fig:pbt_selection}, while the cBathy-derived bathymetry is significantly poorer with the largest error at yFRF = 1200 $\textrm{m}$, which is a consistently recurring issue at the Duck site. The proposed blending method is slightly better than the PBT, since the uncertainty estimates for the cBathy observations are high and there is relatively little correction from cBathy under the given weather conditions. 

\begin{figure}[h!]
    \centering
    \includegraphics[width=0.8\textwidth]{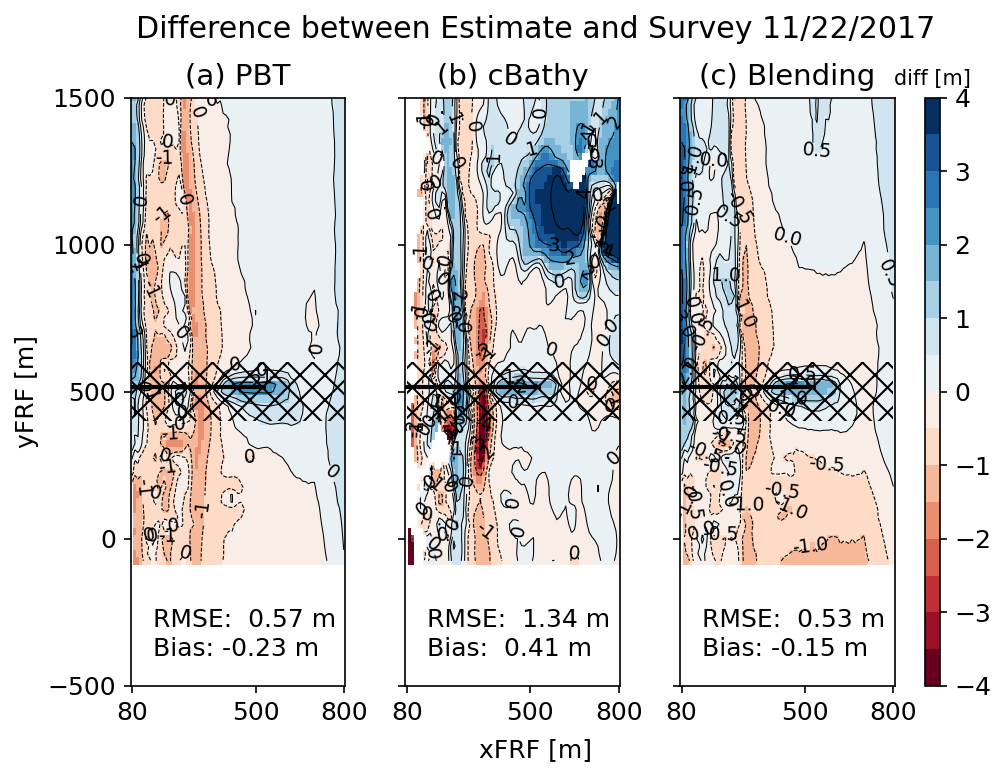}
    \caption{Difference between estimate and survey on 11/22/2017; the estimation error is computed as the difference between the estimate and surveyed bathymetry over the non-pier area of $400$ $\textrm{m}$ $>$ yFRF or yFRF $> 600$ $\textrm{m}$.}
    \label{fig:diff_11222017}
\end{figure}

Figure~\ref{fig:diff_11222017} displays the difference between the survey and each estimate. The identified bathymetry is close to the survey data with RMSE of 0.53 $\textrm{m}$. The RMSE from PBT and cBathy are 0.57 and 1.34 $\textrm{m}$, respectively. The bias of the PBT, cBathy, and blending estimates are -0.23, 0.41, and -0.15 $\textrm{m}$, respectively. As discussed in the previous studies \citep{holman2013cbathy, brodie2018evaluation}, the error of the single hourly data based cBathy estimates relative to the surveys increases with wave height, and the blending estimate also has larger errors than the previously shown blending estimates under mild wave/weather conditions. Even under relative high wave conditions, the blending method shows better results (RMSE = 0.41/Bias = -0.15 $\textrm{m}$). Note that at the Duck site, previously computed Kalman filtered cBathy Phase 3 estimates have shown RMSE and Bias of 0.75 $\textrm{m}$ and -0.25 $\textrm{m}$ on average, respectively, regardless of the wave conditions \citep{holman2013cbathy,brodie2018evaluation} due to the incorporation of multiple time-series bathymetry estimates. Thus, the blending method has a potential to improve the bathymetry estimation further. 

\begin{figure}[h!]
    \centering
    \includegraphics[width=\textwidth]{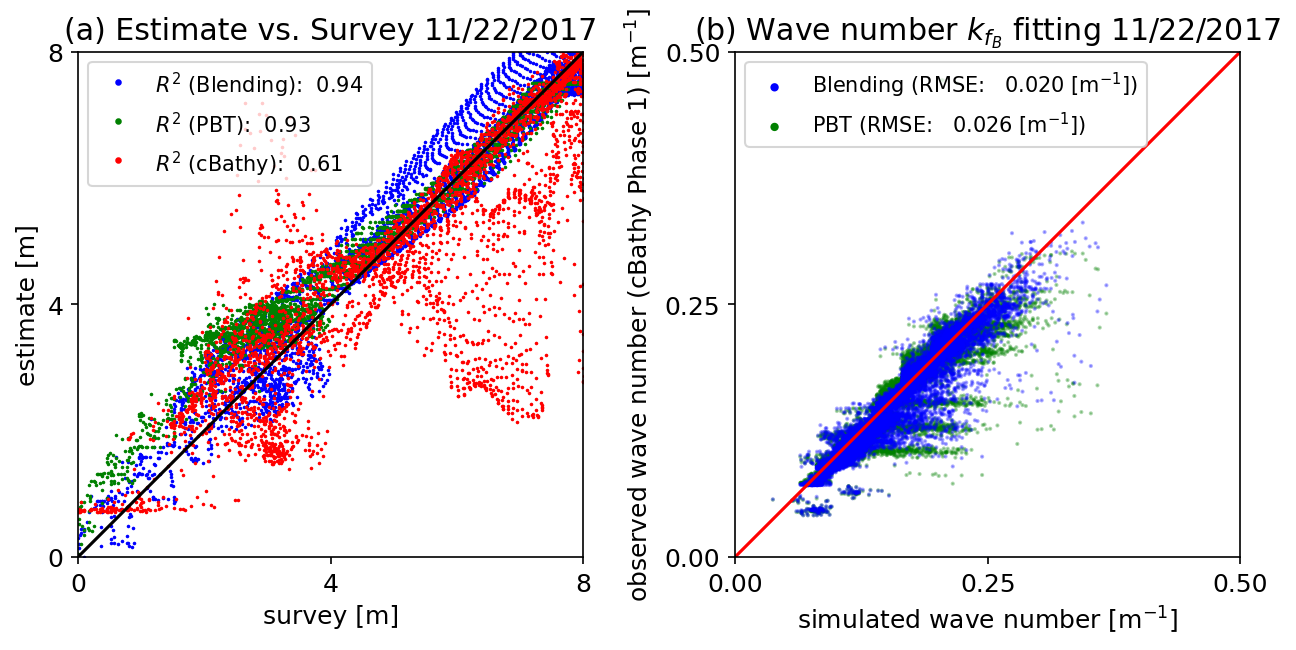}
    \caption{(a) Surveyed vs. estimated water depth, (b) simulated vs. observed wave numbers for PBT and Blending for the 11/22/2017 data set}
    \label{fig:bathy_fit_11222017}
\end{figure}

Figure \ref{fig:bathy_fit_11222017} (a) shows the plot of bathymetry estimation points against survey points with $R^2$ values and confirms the improved $R^2$ values of 0.93 (PBT) and 0.94 (blending) compared to the cBathy $R^2$ value of 0.61. The PBT and blending methods generally overestimate the water depth as found in the computed bias with negative values (-0.25 and -0.15 $\textrm{m}$, respectively).  In Figure~\ref{fig:bathy_fit_11222017} (b), the wave numbers estimated from the blending method and PBT are plotted versus the measured wave numbers. While PBT produces an RMSE of 0.026 $\textrm{m}^{-1}$, the blending method reproduces the wave number more closely with RMSE of 0.02 $\textrm{m}^{-1}$ through the cBathy wave data. 

\begin{figure}[htbp!]
    \centering
    \includegraphics[width=\textwidth]{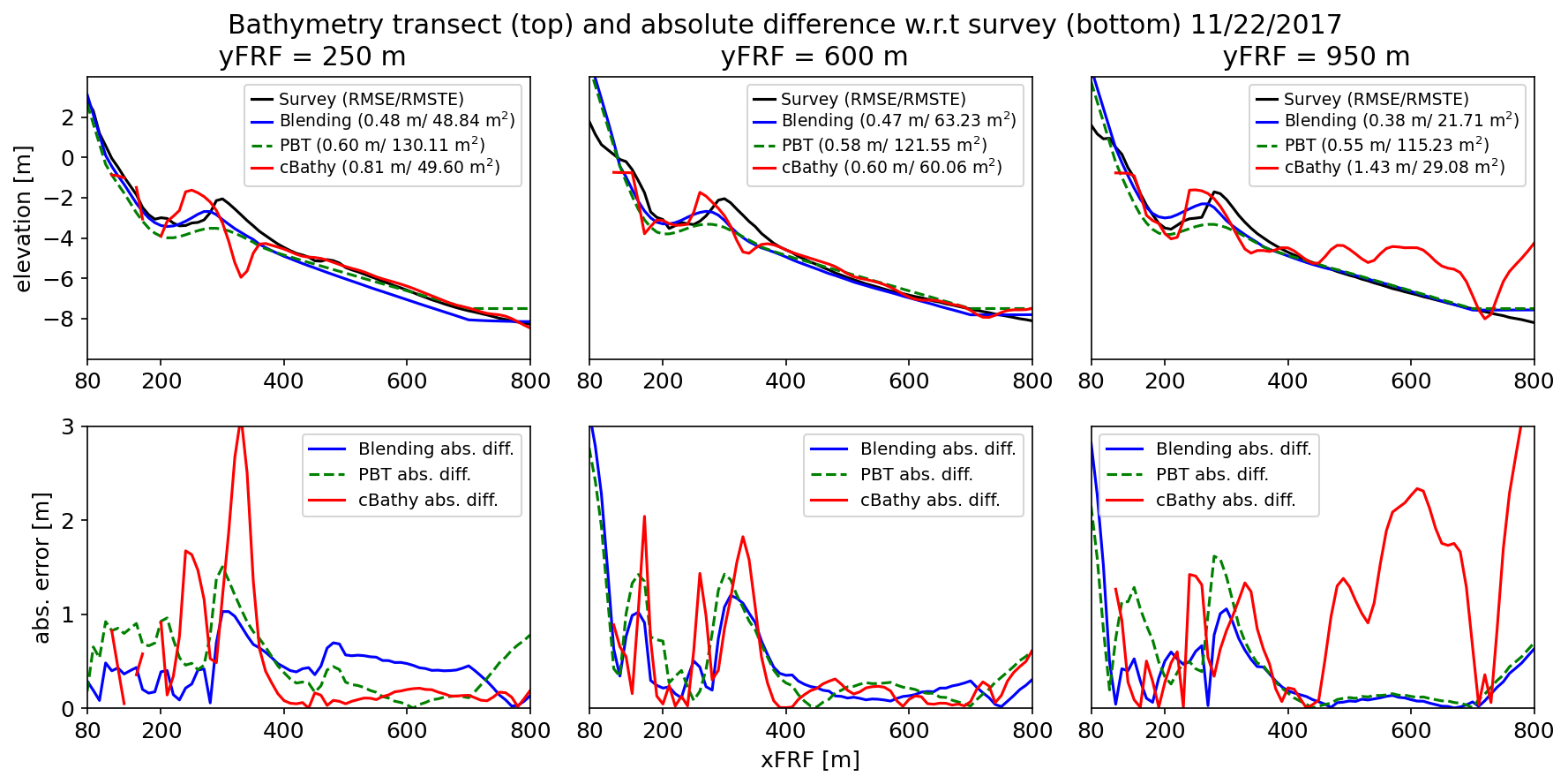}
    \caption{Cross-shore transects (top) and estimation error (bottom) at yFRF = 250, 600, and 950 $\textrm{m}$. Estimation error is computed as difference between estimation and direct survey}
    \label{fig:transect_11222017}
\end{figure}

In Figure ~\ref{fig:transect_11222017}, the estimated bathymetry and survey transects and their absolute errors at yFRF = 250, 600 and 950 $\textrm{m}$ are presented. As in the previous cases, RMSE values obtained from the blending method outperform those from cBathy and PBT consistently for the three transects indicating the superior estimation accuracy of the blending method. The blending method reconstructs the realistic structures of the bar and trough with similar or better RMSTE values compared to the cBathy transect estimates. Note that RMSTE computation is dependent on the boundary conditions of the mass transport (either no transport flux or free boundary) and small difference in RMSTE practically . The results suggest that the blending method can be used for reliable bathymetry identification even under relatively high wave height/wind condition. 

\section{Discussion}
\label{sec:disc}
The results in Section \ref{sec:results} demonstrate that the blending framework can successfully combine cBathy-based observations of wave celerity with a background bathymetry estimate across a range of weather and wave conditions. The error statistics computed from the results for the entire domain and selected transects are summarized in Table \ref{tab:rmsd} and \ref{tab:rmsd_transect}, respectively. 
\begin{table}[h!]
\begin{center}
\aboverulesep=0ex 
\belowrulesep=0ex 
\begin{tabular}{|c|c|c|c|c|c|c|}
\toprule
Date &     \multicolumn{3}{c|}{RMSE [m]}  & \multicolumn{3}{c|}{Bias [m]} \\\cline{2-7}
     & PBT & cBathy & Blending & PBT & cBathy & Blending \\\midrule
3/27/2017  & 0.59 & 0.60 & 0.51 & -0.26 & 0.07 & -0.02  \\\midrule
11/22/2017 & 0.57 & 1.53 & 0.53 & -0.23 & 0.84 & -0.15  \\\midrule
10/6/2019  & 0.47 & 1.29 & 0.42 & -0.17 & -1.00 & -0.23  \\\bottomrule
\end{tabular}
\end{center}
\caption{RMSE and Bias of the estimates for the three cases considered in Section \ref{sec:results}}
\label{tab:rmsd}
\end{table}

\begin{table}[ht!]
\begin{center}
\aboverulesep=0ex 
\belowrulesep=0ex 
\begin{tabular}{|c|c|c|c|c|c|c|c|}
\toprule
Date &   Transect &  \multicolumn{3}{c|}{RMSE [m]}  & \multicolumn{3}{c|}{RMSTE [m$^2$]} \\\cline{3-8}
     & Location & Blending & PBT & cBathy & Blending & PBT & cBathy \\\midrule
3/27/2017 & yFRF = 250 m & 0.47 & 0.71 & 0.55 & 48.05 & 114.63 & 48.36 \\ 
          & yFRF = 600 m & 0.32 & 0.41 & 0.59 & 53.90 & 82.09 & 80.09 \\ 
          & yFRF = 950 m & 0.41 & 0.39 & 0.62 & 30.45 & 30.43 & 32.25  \\\midrule
11/22/2017 & yFRF = 250 m& 0.48 & 0.60 & 0.81 & 48.84 & 130.11 & 49.60  \\
 & yFRF = 600 m& 0.47 & 0.58 & 0.60 & 63.23 & 121.55 & 60.06  \\
 & yFRF = 950 m& 0.38 & 0.55 & 1.43 & 21.71 & 115.23 & 29.08  \\
\midrule
10/6/2019  & yFRF = 250 m & 0.41 & 0.49 & 1.28 & 82.87 & 108.35 & 154.57 \\
  & yFRF = 600 m & 0.48 & 0.51 & 0.91 & 103.56 & 110.47 & 105.62  \\
\bottomrule
\end{tabular}
\end{center}
\caption{RMSE and RMSTE of the transect estimates for the three cases considered in Section \ref{sec:results}}
\label{tab:rmsd_transect}
\end{table}

From one direction, the approach can be seen as a way to condition potentially noisy observations with an informed prior. In this respect, the blending performed quite well. Across the three problems, the blended bathymetry improved RMSE over cBathy Phase 2 by a factor of between 1.2 and 3.1, while the improvements in bias were from a factor of 3.1 to 5.6. Improvements in the RMSTE for the shallow water region of the selected transects on considered during 27 March 2017, 22 November 2017, and 6 October 2019 ranged up to a factor of 1.5, 1.33, and 1.86, respectively. It is worth noting that the identified sandbar and trough in the blending method captures the morphological features much better than those obtained from cBathy Phase 2. 


From the other direction, one can view our approach as a way to improve upon an existing bathymetry estimate through inversion of wave celerity. In this respect, the blending also performed well and improves the estimation in the shallow water region in terms of RTMSE although the relative improvement over the PBT estimate in terms of RMSE was not as dramatic as the improvement over the cBathy Phase 2 results as shown in Tables \ref{tab:rmsd} and \ref{tab:rmsd_transect}. 

The PBT was originally designed for beaches like the FRF, and the actual PBT estimates were carefully selected from high-quality imagery. As a result, the PBT estimates captured the general features of the bathymetry quite well. The major contributions from the blending in many cases was to correct bias from the cBathy Phase 2 derived bathymetry estimation, fill in gaps of known poor estimates, and improve sandbar position and its morphological features.   

The performance of any scheme that attempts to combine multiple bathymetries comes down eventually to how the weights for the contributing bathymetries are selected. Our approach sits within a Bayesian framework and has the advantage that it relies solely on available cBathy Phase 1 quantities to estimate uncertainty in the observations, $\mathbf{C}_{\mathbf{kk}}$, which leads to the optimal weighting to different data sources. As the cBathy degrades in breaking waves \citep{brodie2018evaluation}, the PBT method can fill in the gaps following such weights. The three identified example cases in this work varied wave conditions, and the current method relies heavily on accurate uncertainty estimates from the cBathy Phase 1 diagnostic statistics; \texttt{kErr} of the proposed method will be a function of these uncertainty estimates along with the data screening with \texttt{skill}, \texttt{lambda}, and \texttt{hTempErr}. These uncertainty estimates vary spatially and with frequency and outperformed previous efforts using a scalar error parameter \citep{lee2022icce}. Providing a systematic assessment of the uncertainty in a user-specified prior like the PBT is more difficult. For example, a high-quality background estimate is certainly not guaranteed, and in many cases one may be restricted to simpler qualitative approximations like an equilibrium beach profile or an outdated survey. Here, for the PBT we have selected the log-linear covariance model, the so-called de Wijs model, as a simple, parameter-free approximation \citep{krige1981lognormal}. The log-linear covariance model in 2D is suitable for multi-scale fractal field identification \citep{chiles2012geostatistics}, which results in the flattest (e.g., piece-wise linear interpolation) solution as the best estimate while its uncertainty quantification/realizations can resolve multi-scale features \citep{kitanidis1999generalized}. While we successfully account for the data errors, The current work did not quantify the estimation uncertainty, which will be an area of continued research. Linearized uncertainty, i.e., the standard deviation of the estimation, can be computed with Bayes' rule \citep[e.g.,][]{lee2014large} and requires calibration of an additional hyperparameter, i.e., the ratio between the model error and prior variance \citep{kitanidis1997introduction}. 

Fast execution that is scalable to large datasets was a key factor driving the blending algorithm's design. All of the blending computations presented were performed on Google Cloud via a Google Colab notebook that was equipped with one CPU core and 12 GB RAM. The blending operation took approximately one minute on average showing that near real-time bathymetry estimation is feasible once imagery is processed with cBathy Phase 1. 

The blending algorithm is implemented in a Python package ``blendbathyERDC'' for reproducible research and can be obtained from \url{https://www.github.com/jonghyunharrylee/blendbathyERDC}. Basic use requires two lines of Python, i.e., 1) initialization with cBathy, PBT, and survey file locations, and 2) execution of blending (see Appendix \ref{sec:appendixB}). It also allows users to control the blending process with its parameters such as solver tolerances, maximum GN iterations and so on. Default values are provided, and users do not necessarily need to provide blending/inversion-specific parameters. 

\section{Conclusion}
\label{sec:conc}

In this paper, we have presented and tested new method to estimate improved surf-zone bathymetry based on nearshore imagery collected over a short-term period such as a single flight of UAS. The method uses a simple, parametric form-based bathymetry as an initial guess and then corrects and updates the final estimate with small-scale features by fitting the wave celerity to the linear dispersion relationship. Here, the initial bathymetry and wave celerity were obtained using Parameteric Beach Tool and cBathy (Phase 1), respectively. The method does not need a hyperparameter tuning since cBathy Phase 1 diagnostic statistics are used to determine the measurement error level and data source weighting during the blending. The computational speedup and storage efficiency of the proposed method are achieved through the use of a Conjugate Gradient-Gauss Newton method without dense matrix-matrix multiplications. 

The method was tested against three ground truth surveys at Duck, NC under different weather and wave height conditions. Compared to the cBathy estimates (Phase 2), RMSE and Bias obtained from the proposed method improve by around 50\%. Sandbar shapes and locations are well identified as well. The resulting computational time is approximately one minute for a domain with 6000 cells on Google Cloud via a Google Colab notebook with one CPU, which shows a potential for real-time UAS-based field site characterization. We provide a Python package for reproducible research and a notebook example included in the package will reproduce all the results and figures presented in Results section \ref{sec:results}.

\section{Acknowledgement}
This research was supported in part by an appointment to the Department of Defense (DOD) Research Participation Program administered by the Oak Ridge Institute for Science and Education (ORISE) through an interagency agreement between the U.S. Department of Energy (DOE) and the DOD. ORISE is managed by ORAU under DOE contract number DE-SC0014664. All opinions expressed in this paper are the author's and do not necessarily reflect the policies and views of DOD, DOE, or ORAU/ORISE.


\appendix

\section{Root Mean Squared Transport Error}
\label{sec:appendixA}
One may need evaluation metrics other than pixel-wise error such as RMSE. Root Mean Squared Transport Error (RMSTE) computes a difference between two (bathymetry) fields by considering the volume of sediment needed to transport from one to the other field. In this work, we limit our comparison to 1D RMSTE transect estimates. In general, let $\mathbf{q}$ be a cumulative, depth-integrated transport of constant density sediment from $h_{\textrm{estimate}}$ to $h_{\textrm{survey}}$. The sediment volume balance should be satisfied as: 
\begin{equation}
\nabla \cdot \mathbf{q} = h_{\textrm{estimate}} - h_{\textrm{survey}}
\end{equation}
With this constraint, we find the optimal transport field that minimizes the $L_2$ norm of all possible transport fields:
\begin{equation}
\begin{aligned}
\min_{\mathbf{q}} \quad & \left( \frac{1}{L_{\Omega}} \int_{x\in \Omega} |\mathbf{q}(x)|^2 dx \right)^{1/2}\\
\textrm{subject to} \quad & \nabla \cdot \mathbf{q} = h_{\textrm{estimate}} - h_{\textrm{survey}} \\
\end{aligned}
\end{equation}
where $L_{\Omega}=|\Omega|$ and the RMSTE is given by
\begin{equation}
    \textrm{RMSTE} = \left( \frac{1}{L_{\Omega}} \int_{x\in \Omega} |\mathbf{q}_{min}(x)|^2 d\mathbf{x} \right)^{1/2} 
    \label{eq:rmste}
\end{equation}
By choosing the potential $\phi$ where $\mathbf{q}(x) = \nabla \phi(x)$, the solution to the optimization problem in Eq. \ref{eq:rmste} is equivalent to the solution to the following diffusion problem \citep{bosboom2020optimal}:
\begin{equation}
\nabla^2 \phi_{sol}(x) = h_{\textrm{estimate}} - h_{\textrm{survey}} \nonumber
\end{equation}
and the RMSTE is given by
\begin{equation}
\textrm{RMSTE} = \left( \frac{1}{L_{\Omega}} \int \left(\nabla \phi_{sol}(x)\right)^2 dx \right)^{1/2} = \left( \frac{1}{L_{\Omega}} \int \mathbf{q}_{sol}(x)^2 dx \right)^{1/2} \nonumber     
\end{equation}
An ODE solver \citep{kierzenka2001bvp} is used in this work to evaluate 1D RMSTE over transects and the boundary conditions on the ends are assigned to the closed land boundary ($n\cdot q_{\textrm{land}} = 0$) and free boundary ($\phi_{\textrm{offshore}} = 0$), respectively. 1D RMSTE has units of $\textrm{m}^2$, i.e., area times displacement per unit length. 

\section{Python Class BathyBlendingERDC}
\label{sec:appendixB}
The Python Class ``BathyBlendingERDC'' can be cloned from \url{https://www.github.com/jonghyunharrylee/blendbathyERDC}.

\lstset{language=Python}
\lstset{frame=lines,
    caption={Example use of Class BathyBlendingERDC},
    label={lst:code_direct},
    basicstyle=\footnotesize,
    columns=fullflexible
}
\begin{lstlisting}[language=Python]
import bathyblendingERDC as bathyinv
# 1. Initialize 
params = {'use_testdata':0} # use default test data set (032717 dataset)
DuckBathy = bathyinv.BathyBlending(cbathy_dir='./cbathy', pbt_dir='./pbt', survey_dirr='./survey', params=params)
# 2. Perform blending
blended_bathy = DuckBathy.blend()
# 3. Post processing
DuckBathy.plot_bathy()
DuckBathy.plot_bathy_error()
\end{lstlisting}

\bibliographystyle{unsrtnat}
\bibliography{main}

\end{document}